\documentclass[12pt,emtex]{article}
\usepackage{amsfonts}
\usepackage{amssymb}
\usepackage{amscd}
\newcommand{\BEQ}{\begin{equation}}
\newcommand{\EEQ}{\end{equation}}

\newtheorem{Th}{Theorem}
\newtheorem{lem}{Lemma}
\newtheorem{Lem}{Lemma}
\newtheorem{Prop}{Proposition}
\newtheorem{Cor}{Corollary}

\newtheorem{Ex}{Example}
\newtheorem{Def}{Definition}
\newtheorem{Rem}{Remark}

\def\bea{\begin{eqnarray}}
\def\eea{\end{eqnarray}}

\def\SD{{\Sigma_{D}}}
\def\S{{\Sigma}}
\def\O{{\cal O}}
\def\F{{\cal F}}
\def\C{{\mathbb{ C}}}
\def\CC{{\mathbb{ C}}}

\def\ML{M_\Lambda}
\def\K{\mathcal{K}}
\def\L{\Lambda}
\def\M{\mathcal{M}}
\def\T{\mathcal{T}}

\begin{document}

\begin{titlepage}
\hfill ITEP-TH-76/02 \vskip 2.5cm

\centerline{\LARGE \bf Hitchin system on singular curves I
 }

\vskip 1.5cm \centerline{A. Chervov \footnote{E-mail:
chervov@gate.itep.ru} , D. Talalaev \footnote{E-mail:
talalaev@gate.itep.ru} \footnote{The work of the both authors
has been partially supported by the RFBR grant 01-01-00546,
the work of A.C. was partially supported by the
Russian President's grant 00-15-99296}}

\centerline{\sf Institute for Theoretical and Experimental Physics
\footnote{ITEP, 25 B. Cheremushkinskaya, Moscow, 117259, Russia.}}

\vskip 2.0cm

\centerline{\large \bf Abstract.}

\vskip 1.0cm

In this paper we study Hitchin system on singular curves.
Some examples of such system were first considered by
N. Nekrasov  ( hep-th/9503157 ), but our methods are different.
We consider the curves which can be obtained from the projective line
by gluing several points together or by taking
cusp singularities.
(More general cases of gluing
subschemas will be considered in the next paper).
It appears that on such curves all ingredients  of Hitchin integrable system
(moduli space of vector bundles, dualizing sheaf, Higgs field etc.)
can be explicitly described,
which may deserve independent interest. As a main result we find explicit
formulas for the Hitchin hamiltonians.
We also show how to obtain the Hitchin integrable system on such a
curve as a hamiltonian reduction from a more simple system on
some finite-dimensional space.
In this paper we also work out the case of a degenerate
curve of genus
two and find the analogue of the Narasimhan-Ramanan parameterization
of SL(2)-bundles. We describe the Hitchin system in such coordinates.
As a demonstration of the efficiency of our approach
we also rederive the rational
and trigonometric Calogero systems from the Hitchin system on
cusp and node with a marked point.

\end{titlepage}


\tableofcontents


\newpage
\section{Introduction}
Hitchin system was introduced in \cite{Hit} as an integrable
system on the cotangent bundle of the moduli space
$\mathcal{T}^*\mathcal{M}$ of stable
holomorphic bundles on an algebraic curve $\Sigma$. This phase
space can be obtained by the Hamiltonian reduction by
the gauge group action from the space of
pairs $d_A^{''},\Phi$, where $d_A^{''}$ is the operator defining
the holomorphic structure on the bundle $V$ and $\Phi$ is an
endomorphism of this bundle, more precisely $\Phi \in
\Omega^{0,1}(\Sigma,End(V))$ where the gauge group is the
group of $GL_N$-valued
functions on $\Sigma$. The invariant symplectic structure on the
``big'' space can be written as:
\BEQ
\omega=\int_{\Sigma}Tr \delta \Phi \wedge
\delta d_A^{''}. \label{sym}
\EEQ
The zero level of the moment map
is described by the condition $d_A^{''} \Phi=0$ which means that
$\Phi$ is holomorphic with respect to the induced holomorphic
structure on the bundle $End(V)$. It turns out that the system of
quantities $Tr\Phi^k$, treated as vector functions on the phase
space, Poisson-commute and their number is exactly half the
dimension of the phase space.

The importance of Hitchin system and its generalizations \cite{NN,ER1,Markman}
in modern mathematical physics cannot be overestimated.
Many well-known systems can be obtained as
particular cases. Automatically they inherit the universal construction of
a family of commuting hamiltonians as well as the geometric
description of the hamiltonian flows, the Lax representation, and the
``action-angle'' variables.

This domain is also connected with important questions
in mathematical physics like the
geometric Langlands correspondence \cite{BD,FFR,Fr}, conformal field theory
(in a sense Hitchin system is a
Knizhnik-Zamolodchikov-Bernard equation on the
critical level) \cite{FFR,LO}, non-linear partial differential equations
such as KP \cite{Kr2}, Davey-Stewartson equation \cite{T1},
Nahm's equations describing monopoles \cite{Saksida},
and other problems (see for example \cite{GNR},\cite{DW}).

Despite its importance Hitchin system is far from being fully
investigated.
One of the reasons for such a situation is that the  moduli space
of vector bundles is a complicated manifold and
it is difficult to choose ``good'' coordinates on it to
write down the Hamiltonians explicitly.
Several attempts have been done in \cite{NN} and in \cite{ER11}.
Nevertheless such descriptions appear to be complicated
and do not answer many questions
(at least yet).
So it is important to work out some examples of
Hitchin system which on the one hand are sufficiently simple and
on the other hand are rich enough to find out general methods
for solving Hitchin system
and to understand such phenomena as the separation of variables
and the geometric Langlands correspondence.

The approach elaborated in this paper can be applied in rather specific
cases,
namely when the base algebraic
curve is singular and its normalization is a rational curve.
Its richness is proved by the number of nontrivial examples.
For such curves all ingredients of Hitchin systems
(vector bundles, their endomorphisms, the moduli space of vector bundles,
the dualizing
sheaf, Higgs fields)
can be described very explicitly and in a quite  simple way.
So we  hope that the understanding
of such systems will
shed light on the general case.

We proceed by formulating the main results of this paper.
\subsection{Constructing Hitchin system}
Consider the curve
$\Sigma^{proj}$ which results from gluing $N$ distinct points $P_i$ on $\CC
P^1$ to one point (i.e. the curve which is obtained by adding the smooth
point $\infty$ to the curve $ \Sigma^{aff}= Spec \{ f\in \CC[z]: \forall i,j
~ f(P_i)=f(P_j)  \}$.

\begin{itemize}
  \item A rank $r$ vector bundle on such a curve
corresponds to a rank
$r$  module $\ML$ over the affine part given by the subset of vector-valued
functions $s(z)$ on $\CC$
i.e. $s(z)\in \CC[z]^r$ which satisfy the conditions: $s(P_1)=\L_i
s(P_{i})$. The moduli space of vector bundles on $\Sigma^{proj}$ is the
factor by $GL_r$ of the set of invertible matrices $\L_i,\forall  i=2,...,N$ where
$GL_r$ acts by conjugation.
(See section \ref{vect-bund-sect}, theorem \ref{bund-on-sigm}).
  \item The basis of global
sections of the dualizing sheaf on $\Sigma^{proj}$ can be described as
meromorphic
differentials on $\CC$ given by  $\frac{dz}{z-P_1} - \frac{dz}{z-P_i},
\forall i=2,...,N $ (see  section \ref{DDS}, example \ref{node-k-ex}).
  \item The
endomorphisms of the module $\ML$ are matrix valued polynomials
$\Phi(z)$ such that $\Phi(P_1)=\L_i \Phi(P_i)\L_i^{-1}, \forall i=2,...,N$
(see section \ref{end-aff-sect}, proposition \ref{end-prop}).
The action of $\Phi(z)$ on $s(z)$ is: $s(z)\mapsto \Phi(z)s(z)$.
The space $H^1(End(\ML))$ can be described as the space $gl[z]$ of matrix valued
polynomials factorized by the subspaces:
$End_{out}=\{\chi(z)\in gl[z]
|\chi(z)=const\}$ and $End_{in}=\{\chi(z)\in gl[z]|\chi(P_1)=\L_i \chi(P_i)\L_i^{-1}, \forall
i=2,...,N\}$. The elements of $H^1(End(\ML))$ are the
tangent vectors to $\L_i$,
the element $\chi(z)$ gives the following tangent vector to $\L_i$:
\bea
\delta_{\chi(z)} \L_i = \chi(P_1)\L_i-\L_i\chi(P_i)
\eea
  \item The global sections of $H^0(End(\ML)\otimes \K)$ ("Higgs fields")
are described as
\bea
\Phi (z)=
\frac{ \sum_{i=2,...,N} -\L_i \Phi_i\L_i^{-1} }{z-P_1}dz
+\sum_{i=2,...,N} \frac{\Phi_i}{z-P_i}dz, \label{Lax}
\eea
where $\sum_{i=2,...,N} -\L_i \Phi_i\L_i^{-1}+\Phi_i=0$
(see section \ref{end-k-ssec2}, proposition \ref{hol-dif-k}).
Let us mention that precisely this condition arises as the zero moment
level condition
(see section \ref{1-form-sect}, formula \ref{mom_map}).
The symplectic form on the cotangent bundle to the moduli space
can be described as the reduction of the form on the space $\L_i,\Phi_i,
\forall i=2,...,N$ given by
\bea
- \sum_{i=2,...,N} Tr d(\Phi_i \L_i^{-1})   \wedge d \L_i \label{sf1}
\eea
(see section \ref{1-form-sect}, proposition \ref{prop_form}).
\end{itemize}

{\bf Result 1:} The Hitchin system on the curve $\Sigma^{proj}$
can be described as the system with a phase space which is the hamiltonian
reduction of the space $\L_i,\Phi_i$ with the symplectic form
\ref{sf1}; the
reduction is taken by the group $GL(r)$, which acts by conjugation,
the Lax operator is given by \ref{Lax}.
\vskip 1cm
{\bf Remarks:}
For the case of gluing two points the same Lax operator
has been
proposed by N. Nekrasov (\cite{NN}), though his methods are different from
ours, and the explicit description of bundles, dualizing sheaf,
endomorphisms etc is absent in his approach.
When one glues several groups of points: $P_i=P_j$, $Q_i=Q_j$ ...
it is obvious how to modify all propositions above,
for example the Lax operator becomes:
$$
\frac{ \sum_{i=2,...,N} -\L_i \Phi_i\L_i^{-1} }{z-P_1}dz
+\sum_{i=2,...,N} \frac{\Phi_i}{z-P_i}dz
+
\frac{ \sum_{i=2,...,N} -\tilde \L_i \tilde \Phi_i\tilde\L_i^{-1} }{z-Q_1}dz
+\sum_{i=2,...,\tilde N} \frac{\tilde \Phi_i}{z-Q_i}dz.
$$
Actually one can easily guess the Lax operator above from
the case of gluing two points: one must first consider the gluing
of $N-1$ pairs of points together $P_2=R_2$, $P_3=R_3$..., then take $R_k=P_1$.

Analogously we obtain all propositions for the case of a curve with several
cusps at points $P_i$ on $\CC P^1$.
\begin{itemize}
  \item The curve: $ \Sigma^{aff}= Spec \{ f\in \CC[z]: \forall i
 f'(P_i)=0  \}$.
  \item  The modules: $s(z)\in \CC[z]^r: s'(P_i)=\L_i s(P_{i})$.
  \item The basis of global
sections of the dualizing sheaf:  $\frac{dz}{(z-P_i)^2}$. The
endomorphisms of the module $\ML$:
$\Phi(z)$ satisfying the condition $\Phi'(P_i)=[\L_i, \Phi(P_i)].$
  \item The global sections of $H^0(End(\ML)\otimes \K)$ ("Higgs fields"):
\bea
\Phi (z)=\sum_i (\frac{\Phi_i dz}{(z-P_i)^2} + \frac{[\L_i,\Phi_i]dz}{z-P_i}),
\eea
where $\sum_{i=1,...,N} [\L_i , \Phi_i]=0$.
  \item
The symplectic form:
\bea
- \sum_{i=1,...,N} Tr d\Phi_i    \wedge d \L_i \label{sf2}.
\eea
\end{itemize}

\subsection{Narasimhan-Ramanan parameterization}
It is known since \cite{NR1} that the moduli space of $SL(2)$  vector bundles
on a curve of genus 2 is $\CC P^3$. In this paper we introduce some
analogs of the Narasimhan-Ramanan parameters for the $SL(2)$  vector bundles
on a singular curve of genus 2 - the curve with two cusps at points $P_1,P_2$.
On such curves the vector bundles can be described as bundles corresponding
to the modules $\ML$ defined as $s(z)\in \CC[z]^2: s'(P_i)=\L_i s(P_{i})$,
where $\L_i \in sl(2)$.
Our goal is to express
analogs of the Narasimhan-Ramanan parameters via the $\L_1,\L_2$.
\vskip 0.5cm
{\bf Result 2:} The Narasimhan-Ramanan coordinates over a
 singular curve are:
\bea \tau_1 = Tr \L_1^2 , ~~~ \tau_3 = Tr \L_3^2 , ~~~ \tau_2 = Tr
(\L_1\L_2)+ \tau_1\tau_3\frac{(P_1-P_2)^2}{4}. \eea The Hitchin Hamiltonians in
this case are $$H_1=Tr\Phi_1^2=4p_1^2t_1+p_2^2t_3+4p_1p_2t_2;$$
$$H_2=2Tr\Phi_1\Phi_2+(z_1-z_2)^2Tr[\L_1,\Phi_1]^2$$
$$=4p_1p_2t_1+4p_2p_3t_3+(8p_1p_3+2p_2^2)t_2-2(z_1-z_2)^2
p_2^2(t_1t_3-t_2^2);$$ $$H_3=Tr\Phi_2^2=4p_3^2t_3+p_2^2t_1+4p_2p_3t_2,$$
where $$t_1=Tr\L_1^2,\quad t_2=Tr\L_1\L_2,\quad t_3=Tr\L_2^2$$ and $p_i$ are
the corresponding conjugated variables.
\vskip 0.5cm
This paper is organized as follows: the first section contains all
algebraic-geometric preliminaries. In the second section we work out the
case of a rational curve with double point and cusp and show that the
arising systems are the trigonometric and rational Calogero-Moser
system with spin.
In the third section we treat
the case of a rational curve
with two cusps, which is a curve of algebraic genus 2. We consider the
moduli space of holomorphic $SL_2$-bundles on it and
construct the
analog of the Narasimhan-Ramanan parameterization in the singular case.
In conclusion we state some open problems for future work.

\vskip 1cm
{\bf Acknowledgements.} The authors are grateful for their friends
and colleagues for useful and stimulating discussions:
N. Amburg, Yu. Chernyakov, V. Dolgushev, V. Kisunko, A. Kotov, D. Osipov,
 S. Shadrin, G. Sharygin, A. Zheglov, A. Zotov. The authors are thankful to
B. Machet for careful reading of the manuscript.

\section{Algebraic-geometric background}

\subsection{Curves defined by gluing points with multiplicities.\label{mod}}

Let us consider a curve $\Sigma$ and some effective divisor
$D=\sum_i n_i P_i$ ($n_i>0$) such that $deg D> 1$. One defines a new
curve $\Sigma_D$ by, roughly speaking,  gluing all points $P_i$
with multiplicities $n_i$ to one point $P$; formally speaking we
define the structure sheaf ${\cal O} (\Sigma_D)$ to be a subsheaf
of ${\cal O} (\Sigma)$ with the properties: $f(P_i)=f(P_j);
f^{k}(P_i)=0, k=1,...,n_i-1$. In Serre's terminology this is ``the
curve defined by the module D'' (see \cite{Serre} ch. 4 sect. 4).
The new curve $\Sigma_D$ obviously has one more singular point
$P$, the normalization of $\Sigma_D$ is $\Sigma$ (of course, if
$\Sigma$ is a smooth curve).

{\Ex Main example to keep in mind.}
If we consider $\Sigma=\mathbb{C}^1$ and $D= P_1+P_2$
we obtain the curve $Spec \{ f \in \CC [z]: f(P_1)=f(P_2) \} $
which is called node (or double point in another terminology),
it is an affine curve which can be defined by the equation $y^2=x^2(x+a)$,
$z=\frac{y}{x}, P_1=\sqrt{a}, P_2=-\sqrt{a}$.

{\Ex ~} If we consider $\Sigma=\mathbb{C}^1$ and $D= 2P$ we obtain
the curve $Spec \{ f \in \CC [z]: f^{'}(P)=0 \} $
which is called cusp, it is an
affine curve which can be defined by the equation $y^2=(x-a)^3$,
$z=\frac{y}{x-a}, P=a$.

{\Prop Consider $\CC P^1$ and the effective divisor $D=\sum n_iP_i$ on it.
Consider the curve $\Sigma_D$ which is obtained by
gluing points $P_i$ with multiplicities $n_i$ to one point
as was explained above.
Then the genus (i.e.  $dim H^1({\cal O}_{\Sigma_D})$) of such a curve
equals
$(\sum_i n_i) -1$.
}

The proposition above is obvious: it is enough to cover the curve by
two charts: the first contains the singular point and does not contain
infinity,
the other does not contain the singular point, and to calculate
the \v{C}ech cohomology
of $\cal O$. (On the curve such a covering is acyclic because the chosen
charts are affine manifolds).

{\Rem ~} One sees that the genus of the node and cusp curves
is equal to $1$,
the same as for an elliptic curve. It is not surprising due to
the fact that these curves
are degenerations of  elliptic curves and the genus does not change under
deformation.

\subsection{Canonical (dualizing) sheaf on curves defined by gluing points.
\label{CC}}

\subsubsection{Description of the dualizing sheaf.\label{DDS}}

Recall that the dualizing sheaf on a curve $\Sigma$ is, roughly speaking,
a sheaf $\K$ such that for an arbitrary coherent sheaf $F$ there
is a canonical isomorphism $Hom (F,\K)\cong H^1(F)^{*}$.
On a nonsingular curve the dualizing sheaf is the sheaf of 1-forms,
but for a singular curve the notion of ``1-form'' must be clarified
and the naive definition of the K\"ahler differentials (\cite{Harts}
ch. 2 sect. 8 ) is not the right object.
The general receipt (see \cite{Serre,HM}) for the description
of the dualizing sheaf on a singular curve is the following:
{\Prop
Let $\S_{norm}$ be the normalization of $\S$ and
$\pi:\S_{norm}\rightarrow\S$ the corresponding projection.
The dualizing sheaf $\K$ on the singular curve $\Sigma$ can be described
as the sheaf of meromorphic 1-forms $\alpha$  on $\Sigma_{norm}$
such that $\forall f\in \O(\S)$ and $\forall P\in \Sigma$ it is true that:
\bea \label{hol-dif}
\sum_{P_i \in
 \Sigma_{norm}: \pi(P_i)=P} Res_{P_i} \tilde f \alpha =0,
\eea
 where $\tilde f \in \O(\S_{norm})$
 is the pullback of the function
 $f$ on the singular curve to its normalization, $P_i$ are
points on the
normalization such that they map to the point $P$ on the singular curve
by the normalization map $\pi$.
}

Let us describe explicitly the canonical (dualizing)
sheaf on a singular curve defined by gluing the
points $P_i$ (of the smooth curve $\Sigma$)
with multiplicities
$n_i>0$ to one point. Denote by $D$ the divisor of the
points with multiplicities $D=\sum_i n_i P_i$.

{\Cor The dualizing sheaf ${\K}_{\Sigma_D}$  is defined as
the subsheaf of differential forms ~ $w$ on $\Sigma$
with possible poles in $P_i$ with orders $ord_{P_i} w \ge -n_{i}$
(i.e. the subsheaf of ${\K}_{\S} (D)$ )
with the condition $\sum_{i} Res_{P_i} w=0 $.
}

Our convention is $ord_{0} z^n=n$.
Obviously ${\K}_{\Sigma_D}$ is a coherent sheaf
on $\Sigma_D$.
Moreover one can see directly (or look at \cite{Serre} ch. 4 sect. 11)
that it is a locally free sheaf (i.e. a line bundle on a
singular curve).
The dualizing sheaf is not always locally free. It is true
for complete intersections and arbitrary plane curves (see \cite{Serre} for
the discussion ).

{\Ex ~ \label{node-k-ex} } Consider the node curve $\Sigma$, i.e.,
$\CC P^1 $ with two points $P_1,P_2$
glued together, so  the affine part of this curve is
$Spec \{ f\in \CC[z]: f(P_1)=f(P_2) \}$.
The sections of the sheaf $\K_{node}$ on the chart without infinity
are described as $\frac{cdz}{z-P_1} - \frac{cdz}{z-P_2} +f(z)dz$,
where $f(z)$ is holomorphic.
On the other charts
one obtains the sections of $\K_{node}$ by the usual localization procedure:
on the charts, which
do not contain the singular point, the sections of $\K_{node}$ are
the usual holomorphic $1$-forms.
So the only global section of $\K_{node}$ is $\frac{cdz}{z-P_1} -
\frac{cdz}{z-P_2}.$

One can easily guess
what is going on for the case when we glue $n$ points $P_1,...,P_n$ on $\CC P^1$
together: for example the basis of global holomorphic differentials can
be given by  $\frac{dz}{z-P_1} - \frac{dz}{z-P_i} $, for $i=2,...,n$ .

{\Ex ~ } Consider the cusp curve $\Sigma$, i.e., $\CC P^1 $ with the point $P$
glued with multiplicity 2, so  the affine part of this curve is
$Spec \{ f\in \CC[z]: f'(P)=0 \}$.
The sections of the sheaf $\K_{cusp}$ on the chart without infinity
are described by $\frac{cdz}{(z-P)^2}+ f(z)dz$,
where $f(z)$ is holomorphic. So obviously $\frac{cdz}{(z-P)^2}$
is the only global section of $\K_{cusp}$.

The description of the canonical class on an $n$-cusp curve
when we glue one point $P$ on $\CC P^1$ with
multiplicity $n$ is analogous.
For example the basis of global holomorphic differentials can
be given by  $\frac{dz}{(z-P)^i} $, for $i=2,...,n$.

{\Rem ~ } One can see that the Serre's description of a dualizing sheaf is quite
consistent with the naive arguments for the node and cusp curve. Consider the node
curves: $y^2=x^2(x+a)$ and $z=\frac{y}{x}$ - is the coordinate on the
normalization of this curve. The holomorphic differential on an elliptic curve
is given by the formula: $$\frac{dx}{y}= \frac{dz^2}{z(z^2-a)}=\frac{2dz}{z-\sqrt{a}}-
\frac{2dz}{z+\sqrt{a}},$$ so we obtain a differential which satisfies
Serre's conditions: the orders of the poles are 1 and the sum of its
residues is equal to zero.
If one puts $a=0$ which corresponds to the cusp curve
we obtain as a limit the differential $\frac{2dz}{z^2}.$ It has a
pole of order 2 and residue zero. This is also in accordance with Serre's rule.

\subsubsection{Serre's pairing in the \v{C}ech description of $H^1(F) $
on a singular curve.\label{SD}}

Let us describe the pairing (Serre's duality)
between $H^1({\F})$ and
$H^{0}({\F^* \otimes {\K}_{\SD}})$,
where $\F$ is a flat coherent sheaf on the curve $\SD$.
Recall that $\Sigma$ is a smooth curve and $\SD$ is a singular
curve obtained by gluing the points $P_i$ with multiplicities $n_i$
together, $D=\sum_i n_i P_i$.
In \cite{Serre}, the duality is presented in the most general case
by using the language of distributions (adels). So for the sake on
convenience
we write it down explicitly in our simple case.

For smooth curves, the elements of $H^1({\F})$
in Dolbeault's representation are ``$d\bar z$ forms with values in $\F$''.
The elements of $\K$ can be represented as holomorphic 1-forms,
so the pairing can be given by $\int_{\Sigma} <f,f^*> dzd\bar z$.
For singular curves Dolbeault's approach does not work, at least
naively, so we prefer the \v{C}ech description of $H^1({\F})$, which
works perfectly even for singular curves.

Let us cover the curve $\SD$ by the
two charts $U_P=\SD \backslash \infty$ and
$U_{\infty}=\SD \backslash P$,
where we denote by $\infty$ an arbitrary point in
$\SD$, distinct from $P$ (recall that $P$ is
the only singular point obtained by gluing the points $P_i$ together).
This choice of covering is the most convenient for our
calculation and will be used throughout the paper.
One knows that a curve minus
any point is an affine curve, so this covering is sufficient to calculate
the cohomology of the coherent sheaves: $H^1(\F )=\F(U_{\infty}\bigcap U_{P}) /
\left(\F(U_P)\bigoplus\F(U_{\infty})\right)$.

{\Prop \label{Ser-pair-prop}  The Serre's pairing between
$ f\in H^1({\F})$ and
$h\otimes w \in H^{0}({\F^* \otimes {\K}_{\SD}})$ can be described as follows:
consider $\tilde f \in \F(U_{\infty}\bigcap U_{P})$ the representative
of the element $f$, then the pairing is given by:
\bea \label{SerDual}
<f,h\otimes w>=\sum_i Res_{P_i} <\tilde f, h> w
\eea
This pairing is well-defined (i.e. it does not depend on the choice of the
representative
$\tilde f$) and non-degenerate.
}

{\Cor So one obviously
obtains $$dim H^1(\O_\SD)=dim H^0 (\K_\SD)=H^1(\O_\Sigma)+deg D-1.$$}

Let us sketch why the pairing is well-defined and nondegenerate.
In order to see that the pairing is well-defined one needs to check
that it is zero for $\tilde f \in \F(U_P)$ and for $\tilde f \in
\F(U_{\infty})$.
Indeed,
if $\tilde f \in \F(U_{P})$ then
$g=<\tilde f, h> $ belongs to $\O_\SD(U_P)$ and hence its pullback
$\tilde g$ has the same values at the preimages of the point $P$
$\tilde g(P_i)=\tilde g(P_j)=:g(P)$ and satisfies
the conditions $\tilde g^{(k)}(P_i)=0,k=1,\ldots,n_i-1,~~\forall i.$
Hence:
$$<\tilde f, h\otimes w >=\sum_i Res_{P_i} (\tilde g\otimes w)
= g(P) \sum_i Res_{P_i} w=0,$$
where we used the fact that the sum of the residues of a
meromorphic differential is zero.

For an element $\tilde f\in \F (U_\infty)$ one has
$$<\tilde f,h\otimes w>=\sum_{i} Res_{P_i} (\tilde g\otimes w) =-Res_{\infty}
(\tilde g\otimes w) =0,$$
because both $w$ and $\tilde g$ are regular
at $\infty.$

To show that this paring is nondegenerate it is sufficient to prove that
there is no meromorphic differential $w$ of the prescribed type such that
\begin{equation}\label{ort}
  <f,w>=0\quad \forall f\in \O(U_{\infty}\bigcap U_{P}).
\end{equation}
We can present $n-1$ functions $f_i$ on the normalization
with their only pole of sufficiently high order at $\infty$ and
with nondegenerate matrix
 of their derivatives up to $n_i-1$ order at the points $P_i.$ The
condition of orthogonality $$<f_i,\omega>=0, \quad i=1,\ldots,n-1$$
is a system of linear homogeneous equations on the negative coefficients of
$\omega$ at the points $P_i.$ The only solution of this system is zero vector and
one obtains that the pairing is nondegenerate due to the absence of
holomorphic differentials on $\C P^1.$

\subsection{Holomorphic bundles on singular curves}

\subsubsection{Projective modules over an affine part}
Holomorphic bundles on a non-singular manifold can be described by sheaves
of its sections. Such sheaves are locally free or equivalently (by a general theory)
they are
sheaves of projective modules over the structure sheaf.
The geometric description of a holomorphic
bundle on a singular manifold is problematic in contrast with the algebraic side
which is unambiguous.
{\Def The holomorphic bundle on a singular curve $\S$ is the sheaf
of projective modules over $\O(\S).$}
\\
(It is known that a
projective module is locally free (in Zariski's topology) also for singular
manifolds, so it is equivalent to speak about projective
or locally free modules).
First let us describe the projective modules over the
affine curve $\SD^{aff}$ which is obtained by gluing the
points $P_i$ on $\CC$ with multiplicities $n_i$ to one point.
As usual we denote by $D$ the effective divisor $\sum_i n_iP_i$.
We will describe such modules as submodules of the trivial module
on the normalization.

{\Prop \label{ML-mod}
Consider the curve $\SD^{aff}$ given by $Spec \{ f\in \CC[z]:
\forall i,j ~ f(P_i)=f(P_j); f'(P_i)=...=f^{n_i-1}(P_i)=0 \}$.
Consider the following set of matrices:
invertible matrices
$\L_2,...,\L_{N}$
and arbitrary matrices $\L_i^l \in Mat(r)$, where $i=1,...,N$;
$l=1,...,n_i-1$.  The subset of vector-valued polynomials $s(z)$ on $\CC$ i.e.
$s(z)\in \CC[z]^r$ such that they satisfy the conditions:
$s(P_1)=\L_i s(P_{i})$;
$s^{(l)}(P_i)= \L^i_l s (P_i)$
is a projective module of rank $r$ over the algebra
$\{ f\in \CC[z]:
\forall i,j ~ f(P_i)=f(P_j); f'(P_i)=...=f^{n_i-1}(P_i)=0 \}.$
All projective modules can
be obtained in this way.}
\\
{\bf Notation } Let us denote the module and the
bundle described above by $\ML$. (We will use the
same notation $\ML$ for the vector bundles on
the projectivization $\SD^{proj}$ of the curve $\SD^{aff}$,
we hope that it will not be confusing).

{\Rem ~~} One can easily show (calculating the divisor for example)
that in the
case of rank $r=1$ all these modules are non isomorphic for different
$\Lambda$'s. For the $r>1$ it is certainly not true,
but we will see below that the vector bundles
on the projectivization $\SD^{proj}$ of the curve $\SD^{aff}$
corresponding to these modules are isomorphic iff
all $\Lambda$'s are conjugated by the same constant matrix
$C$.

{\Rem ~~} Let us mention that even in the case $r=1$
if one considers the analytic topology then all bundles
$\ML$ become isomorphic, because from the exponential
sequence one can easily see that $H^1( \O^*)=H^1(\O)$ and
$H^1(\O)=0$ on any affine curve.
But, for projective curves
the GAGA principle guarantees the same  results for
 both the algebraic and analytic setups.
We will be interested in projective curves
so one must not pay too much attention to the remarks above.
\\
{\bf Sketch of Proof.} This proposition is quite simple so let us
only sketch out
its proof. Let $\pi: \CC \to \SD^{aff}$ be the normalization map.
Consider any torsion free module $\F$ of rank $r$.
So $\pi^{*} \F$ is a torsion free rank $r$ module,
but all such modules over $\CC[z]$ are trivial,
so $\pi^{*} \F= \CC[z]^r$.
Consider $\pi_{*}\pi^{*} \F.$ It is isomorphic to
$\CC[z]^r$ considered as a module over $\O(\SD)=\{ f\in \CC[z]:
\forall i,j ~ f(P_i)=f(P_j); f'(P_i)=...=f^{n_i-1}(P_i)=0 \}$,
it is a torsion free, but not a  projective module (the fiber
at the singular point $P$ jumps).
We have the exact sequence $\F \to \pi_{*}\pi^{*} \F \to \CC^{rg}$,
where $\CC^{rg}$ is a skyscraper sheaf at the point $P$,
$r$ is the rank, and $g=deg D-1$.

So we see that any torsion free module $\F$
can be described as the kernel of  the map
$\phi: \CC[z]^r\to ~skyscraper ~ at ~ P$.
Such maps  $\phi$ bijectively correspond to
the maps
$\tilde \phi: fiber~at~ P~of~module~\CC[z]^r \to \CC^{rg}$.
The finite dimensional linear space
$fiber~at~ P~of~module~\CC[z]^r$ in our case
is the space $\oplus_{P_i} \CC^{r(n_i+1)}$.
So in general the kernel
of such a map can be described by the maps:
$\Lambda_i: \CC^{r}_{P_i}\to \CC^{r}_{P_{i+1}}$
and $\Lambda_{i}^{j} : \CC^{r}_{0,P_i}\to \CC^{r}_{j,P_{i}}$.
One can easily see that if we are not in the general
case or if $\Lambda_i$'s are not invertible
the modules will not be projective.
$\blacksquare$

Let us recall that the fiber at a point $P$ of a module $M$
over a ring $R$ is defined as $M^{loc}/ I^{loc} M^{loc} $,
where $I$ is the  maximal ideal of the point $P$
and $loc$ means ``localization at point $P$''.

{\Ex ~ \label{proj-mod-ex1} }
Consider the node (or double point) curve:
$\Sigma= Spec \{ f\in \CC[z] : f(P_1)=f(P_2) \}$.
Then  the rank $1$ modules
(line bundles or rank one torsion free sheaf) are parameterized by one complex
number
$\lambda \in \CC$. They are given by the condition
$ \{ s(z)\in \CC[z] : s(P_1)=\lambda s(P_2) \} $.
Obviously $M_{\Lambda}$ is a
torsion free module. For $\lambda =0$ one can see that it is not a
projective module, because the fiber at the point zero jumps and becomes
two-dimensional,  which is impossible for locally free modules.
It is a nice
exercise to calculate the divisor of the line bundle $\ML$. For $\lambda\neq 0$
one can see that this module is locally free (hence projective).
(A rank $1$ projective module  becomes free on any open set which does not
contain any representative of its divisor).  This example illustrates also that
the moduli space of line bundles (the so called generalized Jacobian) on
a singular curve is non compact. In this case $Jac\cong\C^{*}$ and it is
also an isomorphism of groups, where as usually one considers the
tensor product
as a group operation on line bundles. $Jac$ can be compactified by the torsion
free modules. In this case one should add one module corresponding to
$\lambda =0$, (it coincides with the module $\lambda =\infty$, i.e.  the
module $ \{ s\in \CC[z] : 0= s(P_2) \} $). It can be shown that if one
constructs properly the algebraic structure on the set of torsion free
sheaves of rank $1$ as a manifold it coincides with the curve $\Sigma^{proj}$
itself. This result is related to the fact that the Jacobian
of an elliptic curve is isomorphic to this curve. This can be done by
constructing the Poincar\'e line bundle on the product of the curve with
itself.

Given the divisor $D=P_1+P_2+...+P_N$ one obtains the curve
$\SD= Spec \{ f\in \CC[z] : f(P_1)=f(P_2)
=...=f(P_N) \}$ by gluing the points $P_i\in \CC$ together.
The rank $r$ modules can be described as
subsets of vector-valued polynomials $s(z)$ on $\CC$ i.e.
$s(z)\in \CC[z]^r$ such that they satisfy the conditions:
$s(P_1)=\L_i s(P_{i})$, $i=2,...,N$, where $\L_i$ are arbitrary invertible
matrices.

{\Ex ~ \label{proj-mod-ex2} } Consider the cusp curve:
$\Sigma= Spec \{ f\in \CC[z] : f^{'}(P)=0 \}$,
recall that it means that we glued the point $P$ with
itself with multiplicity $2$. The modules
can be described by $ \{ s(z) \in \CC[z] :
s^{'}(P)=\lambda_1 s(P) \} $.
In this example for all $\lambda_1 \in \CC$ these modules are
projective. So $\CC$ is the moduli space of line bundles. It can
be compactified by adding one point $\lambda_1 = \infty$ (i.e.
the module
$\{ s \in \CC[z] : 0= s(0) \} $, which is the same as the maximal
ideal of the
singular point $z=0$, and the same as the direct image of ${\cal O} _{norm}$
and the same as just $\CC[z]$ considered as a module over our algebra).
A properly introduced algebraic structure will show that this
moduli space is the curve $\Sigma ^{proj}$ itself, not $CP^1$ as one might
think from a naive point of view.

Analogously, for the curve
$\SD= Spec \{ f\in \CC[z] : f^{'}(P)=f^{''}(P)
=...=f^{N}(P)=0 \}$ corresponding to the divisor $D=N P$
the rank $r$ modules can be described as
subsets of the vector-valued polynomials $s(z)$ on $\CC$ i.e.
$s(z)\in \CC[z]^r$ such that they satisfy the conditions:
$s^{(i)}(P)=\L_i s(P)$,
$i=1,...,N$,
where $\L_i$ are arbitrary matrices.

\subsubsection{Vector bundles over the projectivization \label{vect-bund-sect}}
The modules $M_\L$ and $M_{\tilde\L}$ are equivalent, if there exists an
invertible map of modules $K(z):M_\Lambda\rightarrow M_{\tilde\Lambda}$.

Recall that we denoted by $\SD^{proj}$ the projective curve
which we obtain from the affine  curve
$Spec \{ f\in \CC[z] \forall i,j : f(P_i)=f(P_j),
f^{k}(P_i)=0, k=1,...,n_i \}$  by adding one smooth
point at infinity.  The modules $M_{\Lambda}$ give a
vector bundle over
$\Sigma^{proj}$ in an obvious way: we define the sheaf
which is a trivial rank $r$ module over the chart containing infinity
and not containing the singular point and which is the
 module $\ML$ (
or more precisely its localization) over the chart which contains
the singular point.
Let us denote these bundles by $\ML$ (we hope that it will not
be too confusing to denote by the same $\ML$ the projective module
over the affine chart and the corresponding vector bundle on the
projectivization).
The degree of such bundles equals zero.
The vector bundles are equivalent if there exists
an invertible map of modules $K(z):M_\Lambda \mapsto M_{\tilde\Lambda}$
over each chart. So we see that $K(z)$ is a matrix polynomial which
must be regular both at infinity and on the affine part $\SD.$
The only such function $K(z)$ is a constant.
So we obtain:

{\Prop
The vector bundles  $\ML$ over $\Sigma^{proj}$
are isomorphic if there exist
a constant matrix $K$ such that
$\forall i,j~~ \Lambda_i=K \tilde \Lambda_i K^{-1} ;
\Lambda_i^j=K \tilde  \Lambda_i^j K^{-1} $.
}
\\
We obtain the following corollary:
{\Th \label{bund-on-sigm}
The open subset in the space of semistable vector bundles of degree zero
and rank $r$ over the curve $\Sigma^{proj}$
 can be described as
$$\mathcal{M}=\L/{GL_r}$$ where $\L$ is the set of
matrices $\{ \Lambda_i, \Lambda_k^j \} ;
i=2,...,N;k=1,...,N;j=1,...,n_i$, with $\Lambda_i$
invertible and $\Lambda_k^j$ arbitrary, and the $GL_r$-action
is defined by the common conjugation by constant matrices.
}
{\Rem ~} Let us also note that for the bundle $\ML$
the pullback $\pi^* \ML$ is the trivial bundle
on $\CC P^1$, where $\pi: \CC P^1 \to \Sigma^{proj}$ is the normalization map.
Obviously there are lots of bundles $\F$ of degree zero on
$\Sigma^{proj}$ such that $\pi^* \F$ are not trivial bundles but
some bundles of the type $\oplus_{k=1,...r} \O (t_k)$ such
that $\sum t_k=0$.
So by no means we obtain all bundles on
$\Sigma^{proj}$ as bundles $\ML$ for some $\L$.
But nevertheless the general stable and possibly semistable
bundles satisfy the property that
$\pi^* \F $ is a trivial bundle on $\CC P^1$,
and so it is easy to see from our previous description
of projective modules that the general semistable bundles
can be obtained as the bundles $\ML$ for some $\L$.

\subsection{Endomorphisms of $\ML$}

\subsubsection{Endomorphisms of the module $\ML$ over an
 affine chart \label{end-aff-sect}}

In this section we will describe endomorphisms
of the bundles over the curves obtained by gluing
distinct points $P_i$ together and for the cusp
curve; {\it the case of gluing points with multiplicities
is more complicated and will be treated in \cite{CT2} }.

Recall that the module $\ML$ over the algebra $ \{ f\in \CC[z]:
\forall i,j ~ f(P_i)=f(P_j) \}$,
is defined as the subset of the vector-valued polynomials $s(z)$ on $\CC,$ i.e.
$s(z)\in \CC[z]^r,$ which satisfy the conditions:
$s(P_1)=\L_i s(P_{i}), i=2,...,N$.
It is natural to look for endomorphisms of $\ML$
as endomorphisms of $\CC[z]^r$ which preserve the submodule $\ML$.

{\Prop \label{end-prop}
An endomorphism of the module $\ML$ can be described as
a matrix polynomial
$\Phi(z): s(z)\mapsto \Phi(z)s(z),$ which satisfy the condition
\bea
\Phi(P_1)=\Lambda_i\Phi(P_{i})\Lambda_i^{-1} \label{end-cond}
\eea
}
\\
The condition above implies that
$\Phi(z)s(z)$ satisfies: $\Phi(P_1)s(P_1)=\Lambda_i \Phi(P_{i})s(P_{i})$,
so $\Phi(z)s(z)$ is again an element of $\ML$ and
$\Phi(z): s(z)\mapsto \Phi(z)s(z)$ is an endomorphism of $\ML$.

{\Ex ~  \label{end-ex1} } In the abelian case
(i.e. rank 1 modules over any manifold) the condition above is
empty and any element $\Phi(z)$ defines an endomorphism i.e. the sheaf of
endomorphisms of any rank 1 coherent sheaf is just $\cal O$ as in the
regular case.

{\Ex ~  \label{end-ex2} }
Consider the node (or double point) curve $Spec \{ f\in \CC[z],
 f(1)=f(0) \}$. An endomorphisms of the module $\ML$ (which is defined
as $ \{ s\in \CC[z]^r,  s(1)=\Lambda s(0) \}$, for some matrix $\Lambda$)
is given by a matrix-valued polynomial
$\Phi(z)=\Phi_0+\Phi_1 z +\Phi_2 z^2+...$ such that
$\Phi(1)= \Lambda \Phi(0)\Lambda^{-1} $.
Hence $\Phi (z)= \Phi_0 + (\Lambda \Phi_0 \Lambda^{-1} - \Phi_0)z
+z(z-1)\tilde\Phi(z)$, where $\tilde \Phi(z)$ is arbitrary.
When one considers the projectivization of our curve and the bundle
corresponding to $\ML$ on it,
we see that in order to be regular at infinity one
must only consider constant endomorphisms $\Phi(z)=\Phi_0$.
So in order to satisfy the condition $\Phi(1)= \Lambda \Phi(0)\Lambda^{-1} $
one must request
that the matrix $\Phi_0$ commutes with $\Lambda$.
As a corollary we see that there is only $r$-dimensional space
of global endomorphisms for a general module $\ML.$

{\Ex ~  \label{end-ex21} }
Consider the node (or double point) curve $Spec \{ f\in \CC[z],
 f(A)=f(B) \}$. An endomorphism of the module $\ML$
is given by a matrix-valued polynomial
$\Phi(z)=\Phi_0+\Phi_1 z +\Phi_2 z^2+...$ such that
$\Phi(A)= \Lambda \Phi(B)\Lambda^{-1} $.
Hence $$\Phi (z)= \Phi_0 + \Phi_1 z
+(z-A)(z-B)(\tilde \Phi(z)),$$
where $\Phi_0, \Phi_1 $ must satisfy
$\Phi_0 +A \Phi_1= \Lambda ( \Phi_0 + B \Phi_1 )\Lambda^{-1} $ and
$\tilde \Phi(z)$ is arbitrary.
It is more convenient to rewrite this expression as follows:
$$\Phi(z)= \Theta (z-A) - \Lambda \Theta \Lambda^{-1} (z-B)+
(z-A)(z-B)\tilde\Theta(z), $$
where $\Theta$ is an arbitrary constant matrix.
So global endomorphisms are given by  $\Phi(z)= \Theta (B-A),$
with $\Theta$ commuting with $\Lambda$.

{\Ex ~  \label{end-ex22} }
Consider the triple point curve $Spec \{ f\in \CC[z],
 f(P_1)=f(P_2)=f(P_3) \}$. So an endomorphism of module $\ML$
is given by a matrix-valued polynomial
$\Phi(z)=\Phi_0+\Phi_1 z +\Phi_2 z^2+...$ such that
$\Phi(P_1)= \Lambda_2 \Phi(P_2)\Lambda_2^{-1}, \Phi(P_1)=
\Lambda_3 \Phi(P_3)\Lambda_3^{-1}. $
Hence $$\Phi (z)= \Phi(P_1) \frac{ (z-P_2)(z-P_3)}{ (P_1-P_2)(P_1-P_3)} +
\Lambda_2^{-1} \Phi(P_1) \Lambda_2  \frac{ (z-P_1)(z-P_3)}{ (P_2-P_1)(P_2-P_3)}
+$$
$$+
\Lambda_3^{-1} \Phi(P_1) \Lambda_3  \frac{ (z-P_1)(z-P_2)}{ (P_3-P_1)(P_3-P_2)}
+ (z-P_1)(z-P_2)(z-P_3) \tilde \Phi(z).$$
\vskip 1cm
Let us consider the cusp curve
$ Spec \{ f\in \CC[z]:
f'(P)=0; \}$,
recall that the module
$\ML$
is defined as the subset of vector-valued polynomials $s(z)$ on $\CC$ i.e.
$s(z)\in \CC[z]^r$  which satisfy the conditions:
$s'(P)=\L s(P)$.

{\Prop
An endomorphism of the module $\ML$ on a cusp can be
described as a matrix polynomial
$\Phi(z): s(z)\mapsto \Phi(z)s(z)$, which satisfy the condition
\bea
\Phi'(P)=[\Lambda, \Phi(P)] \label{end-cond-cusp}
\eea
}

The condition above is obviously equivalent to
 $(\Phi(P)s(P))^{'}=\Lambda \Phi(P)$ which means that $\Phi$ is really
an endomorphism.

\subsubsection{Endomorphisms of the bundle $\ML$ over the
projectivization}

Consider the projective curve $\SD^{proj}$
which is  obtained by adding one smooth point $\infty$
to the curve
$ Spec \{ f\in \CC[z]:
\forall i,j ~ f(P_i)=f(P_j);  f'(P_i)=...=f^{n_i-1}(P_i)=0   \}$.
An endomorphism of the bundle $\ML$
is given by endomorphisms of the corresponding
modules over each chart.
So an endomorphism of the bundle $\ML$
is an endomorphisms of the module $\ML$ over the affine chart
which is regular at infinity.
In order to be regular at infinity
an endomorphism $\Phi(z)$ must be constant $\Phi(z)=\Phi_0,$
on the other hand an endomorphism must satisfy
the conditions \ref{end-cond}, \ref{end-cond-cusp},
so we see that:

{\Prop
A global endomorphism of the bundle $\ML$
over $\SD^{proj}$
is given by a constant matrix $\Phi_0$
which commute with all $\Lambda_i, \Lambda_i^j$.
}

{\Rem ~}  We see that, if the genus of the curve is greater
than $1$, for the general bundle the endomorphisms
are only scalar matrices; this fact reflects the stability of
general bundles. In the case when the genus equals
one
(node and cusp curves), the general bundle has an $r$-dimensional
space of endomorphisms, which corresponds to the fact that
on genus one curves
the general bundle is a sum of linear bundles.

\subsection{Description of $End(\ML)\otimes K$ \label{end-k-sect}}

\subsubsection{Examples of node and cusp curves}

{\Ex ~ \label{end-k-ex21} }
 Consider the node curve $Spec \{ f\in \CC[z],
f(A)=f(B) \}$.
An endomorphism of the module $\ML$  is given (see example \ref{end-ex21})
by
$$
  \Theta(z)=\Theta (z-A) - \Lambda\Theta\Lambda^{-1} (z-B) +
(z-A)(z-B)\tilde \Theta(z),
$$
where $\Theta, \tilde \Theta(z)$ are arbitrary. The sections of the dualizing module
are given by $$\omega=\frac{c dz}{z-A}+\frac{cdz}{z-B}+holomorphic~in~z.$$
So the sections of $End(\ML)\otimes \K$ can be described as
\begin{equation}\label{exthe}
 \Phi(z)=(B-A)( \frac{ \Lambda \Theta \Lambda^{-1} dz }{z-A} -
\frac{ \Theta dz }{z-B})+holomorphic~in~z.
\end{equation}
Hence the global sections  $H^0(End(\ML)\otimes \K)$ over the projectivization
are $\Phi(z)$'s which are regular at infinity.
The condition that
$\Phi(z)$ has no pole of order greater than $2$ gives that there is no
holomorphic term in the expression (\ref{exthe}).
The condition that the
residue at infinity is zero is equivalent to
$\L \Theta \L^{-1}  -
\Theta=0.$ Hence the global sections are:
$$\Phi(z)=(B-A)( \frac{ \Theta  dz }{z-A} -
\frac{ \Theta dz }{z-B})$$ where $\L \Theta=\Theta \L$. We can see in this
case that $$H^0(End(\ML)\otimes \K)=H^0(End(\ML))\otimes H^0(\K).$$

{\Ex ~ \label{end-k-ex3} }
 Consider the cusp curve $Spec \{ f\in \CC[z],
f'(P)=0 \}$.
An endomorphisms of the module $\ML$  is given
by $$\Theta(z)=\Theta+ [\Lambda_1,\Theta] (z-P)+(z-P)^2 \tilde \Theta(z)$$
where $\Theta, \tilde \Theta(z)$ are arbitrary. The sections of the canonical module
are given by $$\frac{c_1 dz}{(z-P)^2}+c(z)dz$$
where $c(z)$ is holomorphic.
So the sections of $End(\ML)\otimes \K$ can be described as
$$ \Phi(z)=(\Theta + [\Lambda,\Theta ](z-P) ) \frac{dz}{(z-P)^2}
+\mbox{{\it holomorphic in z terms}}$$
The global sections $H^0(End(\ML)\otimes \K)$
are $ \Phi(z)$'s
which are regular at infinity. Hence $[\Theta, \Lambda]=0$,
and the global sections are:
$$\Phi(z)=(B-A) \frac{ \Theta  dz }{(z-P)^2}$$ where $\L \Theta=\Theta \L$.
Here we have the same observation as in the node case
$$H^0(End(\ML)\otimes \K)=H^0(End(\ML))\otimes H^0(\K).$$

\subsubsection{Curves obtained by gluing points without multiplicities
\label{end-k-ssec2}}

Consider the curve $\Sigma^{proj}$ which is the result of gluing $N$ distinct
points on $\C P^1 $, (we consider the case of gluing without multiplicities).
Recall that the affine part of $\Sigma^{proj}$ is given by
$ Spec \{ f(z) \in \CC[z]: \forall i,j ~~ f(P_i)=f(P_j) \}.$
The bundle $\ML$ corresponds to the module $\{ s(z)\in \CC[z]^r:
\forall i=2,...,N ~ s(P_1)=\L_i s(P_i) \}$ over the affine
part.

{\Prop \label{hol-dif-k}
A sections of $End(\ML)\otimes \K$ over
the chart without infinity can be described
as a matrix polynomial:
\bea
\label{hol-dif-k-formula}
\Phi (z)=
\frac{ \sum_{i=2,...,N} -\L_i \Phi_i\L_i^{-1} }{z-P_1}dz
+\sum_{i=2,...,N} \frac{\Phi_i}{z-P_i}dz+
\mbox{holomorphic in z terms},
\eea
where $\Phi_i$ are arbitrary matrices.
\\
The global sections $H^0(\Sigma^{proj}, End(\ML)\otimes \K)$
are described by the formula
\begin{equation}\label{globsec}
  \Phi(z)=\sum_{i=1,...,N} \frac{\Phi_i}{z-P_i}dz
\end{equation}
imposing the conditions
\begin{equation}\label{globcond}
  \sum_{i=1,...,N}\Phi_i=0;\qquad
\sum_{i=2,...,N} -\L_i \Phi_i\L_i^{-1}+\Phi_i=0.
\end{equation}
}
\\
{\bf Proof.}
The claim about the section over the affine chart is trivial. Indeed, over
the affine part
$\Gamma(End(\ML)\otimes\K)=\Gamma(End(\ML))\otimes\Gamma(\K).$ It is the
straightforward consequence of the triviality of $\K$ over the affine chart
even in the algebraic setup. So we can represent the section as
\begin{equation}\label{globrazl}
\Phi(z)=\sum_{i=2,...,N} \Phi_i(z) \otimes \omega_i,
\end{equation}
where
$$\omega_i=\frac{dz}{(z-P_1)(z-P_i)}$$
provide the basis of holomorphic
differentials on $\S^{proj}$ and $\Phi_i(z)$ are the sections
of $End(\ML)$ over the affine part, which means that
$\Phi_i(P_1)=\L_k\Phi_i(P_k)\L_k^{-1}.$ Calculating the residues of the
expression for $\Phi(z)$ one obtains formula (\ref{hol-dif-k-formula})
where $\Phi_i=\Phi_i(P_i).$

To prove the second part of the proposition we must realize that global
sections correspond to sections over the affine chart which can be
continued to regular functions at infinity. The
expression (\ref{hol-dif-k-formula}) is regular at infinity if it
has no term holomorphic in $z$ and if
its residue at infinity is zero. This
imposes the additional condition $$\sum_{i=1,...,N}\Phi_i=0.$$
Conversely,
for every $\{\Phi_i;~i=1,\ldots,N\}$ subject to the conditions (\ref{globcond})
there exist a module $\ML$ endomorphism $\{\Phi_i(z);~i=2,\ldots,N\}$ such that
$$\Phi(z)=\sum_{i=1,...,N} \frac{\Phi_i}{z-P_i}\otimes dz=
\sum_{i=2,...,N} \Phi_i(z) \otimes \omega_i.$$
To prove it let us take local endomorphisms $\Phi_i'(z)$ such that
$\Phi_i'(P_i)=(P_i-P_1)\Phi_i ~~i=2,\ldots,N.$ The residue at $P_1$ of
$$\Phi'(z)=\sum_{i=2,...,N} \Phi_i'(z) \otimes \omega_i$$ is
$Res_{P_1}\Phi'=-\sum_{i=2}^N\L_i\Phi_i\L_i^{-1}=\Phi_1.$ The difference
$\Phi(z)-\Phi(z)'$ is regular at the affine chart so it has no
 residue at $\infty.$
Consider now the expression
$$\tilde\Phi(z)\otimes dz=\prod_{i=1}^N(z-P_i)\tilde\Phi(z)\otimes
\frac {dz}{\prod_{i=1}^N(z-P_i)}$$
which is also the section of $End(\ML)\otimes \K$ over the affine chart for
an arbitrary matrix function $\tilde\Phi(z).$ It has arbitrary poles at
$\infty$ of order greater then $2$ and has no residue. One can find
a function $\tilde\Phi(z)$ such that $\Phi(z)=\Phi'(z)+\tilde\Phi(z)\otimes
dz$ but the latter is also an expression of the type (\ref{globrazl}),
which ends the proof $\blacksquare$

\subsection{Description of $H^1(End(\ML))$ \label{h1-end-sect}}

\subsubsection{Curves obtained by gluing points without multiplicities}

Consider the curve $\Sigma^{proj}$ which is the result of gluing $N$ distinct
points on $\C P^1,$ (we consider the case of gluing without multiplicities).
Recall that the affine part of $\Sigma^{proj}$ is given by
$ Spec \{ f(z) \in \CC[z]: \forall i,j~  f(P_i)=f(P_j) \}$,
the bundle $\ML$ corresponds to the module which,  over the affine
part, is described as $\{ s(z)\in \CC[z]^r: \forall i=2,...,N ~ s(P_1)=\L_i s(P_i)
\}$.

{\Prop
\label{lem-chi}
The space of matrix polynomials
$\chi(z)=\sum_{i=0}^{N-1} \chi_i z^i$ maps surjectively to $H^{1}(End(\ML))$.
The kernel of this map is the sum of two linear subspaces
in the space of matrix polynomials $\chi(z)$:
the first space is the space of constant polynomials $\chi(z)=\chi_0$
and the second space is made of matrix polynomials
which satisfy the condition:
$\chi(P_1)=\Lambda_i \chi(P_i)\Lambda_i^{-1}$, for $i=2,...,N$.
(Let us mention that the intersection of these two subspaces is
precisely $H^{0}(End (\ML))$
and that the second subspace consists of $\chi(z)$ which
gives endomorphism of the module $\ML$ over the affine chart without
infinity.)
}
\\
{\bf Proof.} The proposition is quite obvious from the
point of view of the \v{C}ech's description of
$H^1(End(\ML))$. Let us cover our singular curve
by the charts $U_{P} = \Sigma \backslash \infty$,
$U_{\infty}=\Sigma \backslash P,$ where $P$ is the singular point.
Then $H^1(End(\ML))= End(\ML) (  U_{P} \bigcap U_{\infty})
/  End(\ML) (  U_{P})  \oplus End(\ML) (U_{\infty})$.
As we know from  proposition~\ref{end-prop}
$End(\ML) (  U_{P}) $ are precisely polynomials
$\chi(z)$ which satisfy
$\chi(P_1)=\Lambda_i\chi(P_i)
\Lambda_i^{-1}$.  So we obviously come
to the desired conclusion $\blacksquare$

{\Rem~}From the  proposition above we see that
the Riemann-Roch theorem for the bundle $\ML$
becomes obvious. The description of $H^1(End(\ML))$ given above
shares similarities with the description done in
Krichever's and adelic approaches (see e.g. \cite{Denis}).
These descriptions hypothetically
can be used for the proof of the general Riemann-Roch theorem.

{\Ex~\label{ex-tang-lamb-0}}In the abelian case (i.e. when $\ML$ is a
rank
$1$ module) for any $\Lambda$ it is known that $\ML$ is just $\mathcal{O}$.
So in the abelian case the proposition claims that
$H^1(\mathcal{O})$ is the factor space of the space of all polynomials
$\sum_{i=0}^{N-1} \chi_i z^i$ by the space of polynomials
$\chi(z)$ which satisfy the conditions $\chi(P_1)=\chi(P_i)$.
For example this can be seen from the
exact sequence: ${\mathcal{O}}\to {\mathcal{O}}^{norm}\to {\CC_{P}}$
which gives: $H^1(\mathcal{O})=H^0(\ {\CC_{P}} )$.

It is well-known that the vector space
$H^{1}(End(\ML))$ is the tangent space to deformations
of $\ML$ as an algebraic vector bundle, on the other hand we know that all
vector bundles are given by
$\Lambda.$ Our goal is to determine
$\Delta_{\Lambda}$ corresponding to the element $\chi(z)=\chi_i z^i.$

{\Prop \label{chi-deform-lamb}
The matrix polynomial $\chi(z)=\sum_{i=0,...,N-1} \chi_i  z^i$,
(which is considered as an element
of $H^{1}(End(\ML))$ due to the proposition above), gives the following
deformation of $\Lambda_i$:
\bea
\delta_{\chi(z)} \Lambda_i=
\chi(P_1)\Lambda_i- \Lambda_i\chi(P_i).
\eea
}

{\Rem~}One knows that $H^{2}(Coherent ~  sheaves)=0$ for the case
of curves and so by the general theory
the map from $H^1(End(\ML))$ to the tangent space of deformations of
the bundle $\ML$ is a bijection. So the formula above can be taken
as a definition
of the map from the space of matrix polynomials $\chi(z)=\sum_i \chi_i z^i$
to the space $H^1(End(\ML))$. It means that we can
(by definition) associate with the matrix polynomial $\chi(z)=\sum_i \chi_i z^i$
an element of $H^1(End(\ML))$ which deforms the bundle $\ML$
by the formula
$\Lambda_i \mapsto \Lambda_i +
\chi(P_1)\Lambda_i - \Lambda_i \chi(P_i)$.
What must be  proved after such a definition is how to describe
the Serre's pairing between $H^0(End (\ML) \otimes K)$ and
$H^1(End (\ML))$.
We describe Serre's pairing in  proposition
\ref{lem-serre-pair}.

{\Cor
The matrix polynomial $\chi(z)=\sum_j \chi_j z^j$,
which for $i=2,...,N$ satisfies the condition
$\chi(P_1)=\Lambda_i\chi(P_i)\Lambda_i^{-1}$
does not change $\Lambda_i$. This fact is in full
agreement with proposition \ref{lem-chi} which
says that such a polynomial gives a zero element
in $H^1(End(\ML))$.
}
{\Cor
The matrix polynomial $\chi(z)=\chi_0$,
changes $\Lambda$ to its conjugated by a constant matrix, so
it gives the same vector bundle. This fact is in
full agreement with proposition \ref{lem-chi} which says that such a
polynomial
gives a zero element in $H^1(End(\ML))$.  }
\\
{\bf Proof.}
This proposition can be demonstrated as follows:
consider an element $ 1+\delta\chi(z)$, where $\delta^2=0$,
it is an infinitesimal automorphism of the module $\CC[z]^{r}$ corresponding
to the endomorphism $\chi(z)$.
Having such an automorphism it is clear how to deform the
module $\ML$:  the new module is the set of elements
$(1+\delta\chi(z)) s(z)$, where $s(z)$ is an element of $\ML$.
The elements of the type $\tilde s(z)=(1+\delta\chi(z)) s(z)$
satisfy the condition:
$$\tilde s(P_1)= (1+\delta\chi(P_1))
\Lambda_i(1+\delta\chi(P_i))^{-1}
\tilde s(P_i)$$ for $i=2,...,N.$ Hence
$$\tilde s(P_1)= (\Lambda_i+\delta\chi(P_1)\Lambda_i-
\Lambda_i\delta\chi(P_i)) \tilde s(P_i).$$
We see that the new module
is the  module $M_{\Lambda_i +
\chi(P_1)\Lambda_i - \Lambda_i \chi(P_i)}$. $\blacksquare$

Our reasoning
has been the following: the module $\ML$
is embedded in the module $\CC[z]^r$, this module cannot be deformed,
so the deformations of $\ML$ are governed only by the deformations of the
embedding $\ML \to \CC[z]^r$. The elements of $H^1(End (\ML))$
have been identified with some elements of $End(\CC[z]^r)$
due to the fact that  $\CC[z]^r\to \CC^p$ is a resolution
of $\ML$. The elements of $End(\CC[z]^r)$ act on the embeddings
of $\ML\to \CC[z]^r $.
But the formulas in the paragraph above
are more convincing than any words.

There exists another way to demonstrate the proposition above.
It is more transparent at the level of ideas but much
longer at the level of formulas. Let us use the \v{C}ech's description
of cohomologies of sheaves. We consider the covering
of our projective curve consisting of two charts:
the first is everything except the singular point, the second
is not really an open set but a limit
of the open sets - infinitesimal neighborhood of the singular point
i.e. $Spec \{ f\in
\CC(z), f$ is regular at $P_i$ and  $f(P_i)=f(P_j) \}$.
It is convenient to consider such an infinitesimal neighborhood.
Only in such a neighborhood of the singular point all modules
become trivial because it is the spectrum of a local ring.
So any module on a singular curve can be given by gluing
two trivial modules by the gluing function on the intersection
of two charts. In this case the intersection is the ``general point''
i.e. $Spec \{ \CC(z)\}$.
So the first task is to describe the module $\ML$ by the
gluing function.
After that it is obvious how to calculate which
deformation corresponds to an element of $H^{1}(End(\ML)).$
We represent an element of $H^{1}(End(\ML))$ as an element
$\chi\in End(\ML)$ on the intersection of the two charts and
one must simply multiply the gluing function by the
element $1+\delta \chi$. So we obtain a new gluing function. The
new bundle can be again represented in the form $\ML$,
so we obtain the deformation of $\ML$ and this construction
gives the same results as above.

Let us give examples  illustrating
propositions \ref{lem-chi} and  \ref{chi-deform-lamb}.

{\Ex ~ \label{ex-tang-lamb-1}}
Consider the node curve with the affine part
$Spec \{ f(A)= f(B) \}$, where $A,B \in \CC$.
Consider the matrix polynomial
$$\chi(z)=\chi_0+\chi_1 z + (z-A)(z-B)\tilde \chi(z).$$
According to proposition \ref{chi-deform-lamb}
it acts on $\Lambda$ by the formula
$\delta_{\chi}\Lambda = \chi(A)\Lambda-\Lambda \chi(B).$
The part $(z-A)(z-B)\tilde \chi(z)$
does not act on $\Lambda.$
So we can consider only the linear part $\chi(z)=\chi_0+\chi_1 z$.
According to proposition \ref{lem-chi} the
$H^1(End(\ML))$ is the factor of the space $\chi_0+\chi_1 z$
by the sum of the spaces $\chi(z)=\chi_0$ and $\chi(z)=\Theta(z-A)-
\Lambda \Theta \Lambda ^{-1}(z-B), $ where $\chi_0$, $\Theta$ are
arbitrary matrices. (It would be nice
to have an explicit parameterization of the orthogonal
complement with respect to the  Killing form
to the sum of these two  subspaces and to generalize it to the case of
schematic points). The intersection of these two subspaces
is the subspace  of $\chi(z)=\chi_0$ such that $\chi_0$
commutes with $\Lambda.$ This intersection is
$H^0(End(\ML))$. So we see that
$dim H^1 (End(\ML))= dim H^0(End(\ML))$.
This observation coincides with the calculation done from the Riemann-Roch
theorem: $$dim H^0 (End(\ML))- dim H^1(End(\ML)= deg (End (\ML))
- n^2 (1-dim H^1( \mathcal{O)})=0.$$

{\Prop \label{lem-serre-pair}
The Serre's pairing between $H^{0}(End(\ML)\otimes \K)$ and
$H^{1}(End(\ML))$
can be written in terms of the
matrix polynomials  $\tilde \Phi(z)dz$
and  $\chi(z)$ as follows:
\bea
{\label{pairing-1-0}}
\sum_{i=1,...,N}  Rez_{P_i } Tr \chi(z) \tilde \Phi(z) dz.
\eea
}

{\Cor Consider the matrix polynomial $\chi(z)=\sum_k \chi_k z^k$
which satisfies the conditions
$\chi(P_1)=\Lambda_i \chi(P_i) \Lambda_i^{-1}$ for $i=2,...,N$.
The Serre's pairing (given by formula
\ref{pairing-1-0})  between $\chi(z)$ and an
arbitrary $\Phi(z)\in H^{0}(End(\ML)\otimes \K)$
is identically zero.
This fact is in full
agreement with proposition \ref{lem-chi} which
says that such a polynomial gives a zero element
in $H^1(End(\ML))$.
}

{\Cor
Consider the matrix polynomial $\chi(z)=\chi_0.$
The Serre's pairing given by formula
\ref{pairing-1-0} between $\chi(z)$ and an
arbitrary $\Phi(z)\in H^{0}(End(\ML)\otimes \K)$
 is identically zero.
This fact is in full agreement with proposition
\ref{lem-chi} which says that such a polynomial
gives a zero element in $H^1(End(\ML))$.  }

This proposition follows immediately from the general
description of Serre's pairing given in proposition
\ref{Ser-pair-prop}.
To prove the corollaries we use  proposition
\ref{hol-dif-k} in order to  represent $\tilde \Phi(z)$ as
a matrix polynomial:
$$\frac{ \sum_{i=2,...,N} -\L_i \Phi_i\L_i^{-1} }{z-P_1}dz
+\sum_{i=2,...,N} \frac{\Phi_i}{z-P_i}dz$$
for some $\Phi_i.$

{\Rem ~}
To prove the second corollary we must  also use the condition
$$\sum_{i=2,...,N} -\L_i\Phi_i\L_i^{-1}  + \Phi_i =0.$$
The first corollary does not use this condition for $\Phi_i$
and is true for all $\Phi(z)$ represented in the form
$$\Phi(z)=\frac{ \sum_{i=2,...,N} -\L_i \Phi_i\L_i^{-1} }{z-P_1}dz
+\sum_{i=2,...,N} \frac{\Phi_i}{z-P_i}dz$$
with arbitrary $\Phi_i$.

\subsubsection{The cusp curve}

Consider the cusp curve $\Sigma^{proj}$.
Recall that the affine part of $\Sigma^{proj}$ is given by
$ Spec \{ f(z) \in \CC[z]: f'(P)=0 \}$,
the bundle $\ML$ corresponds to the module which is described as
$\{ s(z)\in \CC[z]^r: s'(P)=\L s(P)\}$ over the affine part.

{\Prop
\label{lem-chi-cusp}
The space of matrix polynomials
$\chi(z)= \chi_0+\chi_1 (z-P)$ maps surjectively to $H^{1}(End(\ML))$.
The kernel of this map consists of the sum of two linear subspaces
in the space of matrix polynomials $\chi(z)$:
the first space is the space of constant polynomials $\chi(z)=\chi_0$
and the second space consists of matrix polynomials
which satisfy the condition:
$\chi_1=[\Lambda, \chi_0]$ for $i=2,...,N$.
(Let us mention that the intersection of the two subspaces is
precisely $H^{0}(End (\ML))$ and the second subspace consists of $\chi(z)$ which
gives endomorphisms of the module $\ML$ over the affine chart without infinity.)
}
\\
{\bf Proof. } This proposition is quite obvious and can be
demonstrated
in the \v{C}ech approach in the same way as proposition \ref{lem-chi}.

{\Prop \label{chi-deform-lamb-cusp}
The matrix polynomial $\chi(z)= \chi_0+\chi_1 (z-P)$ which is
considered as an element
of $H^{1}(End(\ML))$ due to the proposition above gives the following
deformation of $\Lambda$:
\bea
\delta_{\chi(z)} \Lambda= \chi_1+[\chi_0,\L].
\eea
}
\\
{\bf Proof. } This proposition can be demonstrated
 in the same way as  proposition \ref{chi-deform-lamb}.
Let us only comment on the key step.
Consider an element $ 1+\delta\chi(z)$, where $\delta^2=0$,
it is an infinitesimal automorphism of the module $\CC[z]^{r}$ corresponding
to the endomorphism $\chi(z)$.
Having such an automorphism it is clear how to deform the
module $\ML$:  the new module is the set of elements
$(1+\delta\chi(z)) s(z)$ where $s(z)$ is an element of $\ML$.
The elements of the type $\tilde s(z)=(1+\delta\chi(z)) s(z)$
satisfy the condition:
$$\tilde s'(P)= (1+\delta\chi(P))' s(P)+ (1+\delta\chi(P)) s'(P)
$$
$$=\delta\chi(P)'(1-\delta\chi(P)) \tilde s(P)+
(1+\delta\chi(P)) \L (1-\delta\chi(P)) \tilde s(P)$$
$$
=(\L+\delta\chi(P)'+[\delta\chi(P),\L])\tilde s(P).$$
Hence one can see that the new module
is the module $$M_{\L+\delta\chi(P)'+[\delta\chi(P),\L]}.$$ $\blacksquare$
\\
The Serre's pairing can be described exactly in the same way as
in subsection above.

\subsection{Canonical 1-form on the cotangent bundle to the moduli space of
vector bundles in terms of $\Phi,\L$ \label{1-form-sect}}

\subsubsection{Curves obtained by gluing points without multiplicities }

Consider the curve $\Sigma^{proj}$ which is result of gluing $N$ distinct
points on $\C P^1 $ without multiplicities.
Recall that the affine part of $\Sigma^{proj}$ is given by
$ Spec \{ f(z) \in \CC[z]: \forall i,j~~  f(P_i)=f(P_j) \},$
the bundle $\ML$ over the affine
part corresponds to the module $\{ s(z)\in \CC[z]^r:
\forall i=2,...,N ~ s(P_1)=\L_i s(P_i)
\}$.

It is well-known that $H^1(End(\ML))$ is the tangent space to the moduli space
of  vector bundles at the point $\ML$ and  $H^0(End(\ML)\otimes \K)$ is the dual
space to $H^1(End(\ML))$ so it is the cotangent space to
the moduli space of vector bundles at the same point.
According to proposition \ref{hol-dif-k} the sections of
$H^0(End(\ML)\otimes \K)$ can be described as:
\bea
\label{hol-dif-k-formula-2}
\Phi (z)=
\frac{ \sum_{i=2,...,N} \L_i \Phi_i\L_i^{-1} }{z-P_1}dz
-\sum_{i=2,...,N} \frac{\Phi_i}{z-P_i}dz
\eea
where the matrices $\Phi_i$  satisfy
$\sum_{i=2,...,N} -\L_i \Phi_i\L_i^{-1}+\Phi_i=0$.

This section is devoted to proving the claim which, in expert
language, can be formulated as follows:
\\
{\bf Claim:}
{\it
The canonical 1-form on the cotangent bundle to the
moduli space of vector bundles on the curve $\Sigma^{proj}$
in terms of $\L_i,\Phi_i$ can be written in the form:
\bea \label{1-form}
- \sum_{i=2,...,N} Tr \Phi_i \L_i^{-1} d \L_i
\eea}
{\Rem ~}
In the abelian case (the case of moduli space of line bundles)
the claim above is an exact proposition -
$\L_i, \Phi_i$ are  ``honest'' coordinates on the cotangent to the
moduli space of line bundles, and the expression above has a clear
meaning.
In the nonabelian case
there is a subtlety due to the fact that the moduli space of vector bundles
is the factor of the space of matrices $\L_i$ by conjugation,
but the $1$-form above is written
on the space of $\L_i, \Phi_i$ without factorization. So one must
explain what the expression $\Phi_i \L_i^{-1} d \L_i$ means,
if $\L_i$ are defined only up to conjugation.
It is obvious for the expert, nevertheless,
despite it is a bit long. Let us give a correct formulation of the proposition,
do not claiming that everybody can translate from
the expert's slang to the precise formulation.

Consider the space of matrices $\L_i, \Phi_i$ with the $1$-form
$\sum_i \Phi_i\L_i^{-1}d \L_i$. Consider the subspace defined by the
equation $\sum_{i=2}^N(\L_i\Phi_i\L_i^{-1}-\Phi_i)=0$.
Consider the map $p$ of this subspace  to the cotangent to the moduli space
of bundles given by $p: (\L_i, \Phi_i) \mapsto (\ML, \Phi(z))$,
where $\Phi(z)$ is defined by formula (\ref{hol-dif-k-formula-2}).
Let us define the $1$-form on the cotangent to the moduli space
as follows: restrict the 1-form
$\sum_i \Phi_i\L_i^{-1}d \L_i$ to the submanifold
$\sum_{i=2}^N(\L_i\Phi_i\L_i^{-1}-\Phi_i)=0$; check that the $1$-form equals zero
on the tangent vectors to the fiber of the map $p$;
hence the $1$-form $\sum_i \Phi_i\L_i^{-1}d \L_i$ can be
pushed down to the image of $p$ i.e. to
the  cotangent to the moduli space
of vector bundles. We claim that the {\it result of the push-down of
$\sum_i \Phi_i\L_i^{-1}d \L_i$ under the map $p$ coincides with
the canonical $1$-form on the cotangent bundle to the moduli space. }
The procedure above is actually a hamiltonian reduction.
Now we proceed with formulating the exact result.

Let us consider the cotangent space to the space of matrices $\L_i$ which is
$$\T^*(GL_r^{\times (N-1)})=\times_{i=2}^N\T^*(GL_r)$$ with the canonical
invariant symplectic form on it $$\omega_1=\sum_{i=2}^N
Trd(\Phi_i\L_i^{-1})\wedge d(\L_i),$$
where $\Phi_i$ are coordinates on $\T^*(GL_r^{\times (N-1)})$
that  $\Phi_i\L_i^{-1}$ are coordinates on $\T^*(GL_r^{\times (N-1)})$
canonically conjugated to the coordinates
$\L_i$ on $(GL_r^{\times (N-1)}).$
(When we say that the matrix $M$ is a ``coordinate''
we mean that each of its matrix elements is a coordinate).
This symplectic form is invariant by the natural action of
$GL_r$:
$$g:\L_i\mapsto g^{-1}\L_i g;\qquad\Phi_i\mapsto g^{-1}\Phi_i g.$$
The moment map of this action $\mu: \T^*(GL_r^{\times (N-1)}) \to gl_r^*$
can be calculated by using the $1$-form
$\lambda_1=\sum_{i=2}^N Tr\Phi_i\L_i^{-1}d(\L_i)$ due to
its invariance as follows:
$$(i(\xi)\lambda_1)[P] =Tr(\mu[P]*\xi)$$
where $P\in\T^*(GL_r^{\times (N-1)})$ and we use
the identification of the vector spaces $gl_r^*$ and  $gl_r$ by means
of $<A,B>=Tr AB.$
Here $\xi$ is the vector field
corresponding to the infinitesimal action of $gl_r$
which can be written as:
$i(\xi)d\L_i=\xi \L_i-\L_i\xi.$
Finally one obtains $$i(\xi)\lambda_1=\sum_{i=2}^N
Tr\Phi_i\L_i^{-1} (\xi \L_i-\L_i\xi)$$ such that the moment map as an
element of $gl_r^*$ coincides with
\begin{equation}\label{mom_map}
  \mu[\L_i,\Phi_i]=\sum_{i=2}^N(\L_i\Phi_i\L_i^{-1}-\Phi_i).
\end{equation}

{\Rem~} We see that the holomorphity condition for $\Phi(z)$ (see proposition
\ref{hol-dif-k-formula}) coincides with the condition that the moment map equals
zero.

Let us consider the hamiltonian reduction for the action above corresponding
to the zero moment level
$$\T^*(GL_r^{\times (N-1)})//{GL_r}=\mu^{-1}(0)/GL_r.$$
This space is endowed with the canonical symplectic structure which
coincides
with the symplectic structure on the cotangent bundle
$$\T^*(GL_r^{\times (N-1)}/{GL_r}).$$
For pedagogical reasons we recall here the proof of the following well known
result:
{\Lem Let the group $G$ act on the manifold $M$; then there is an induced
action of $G$ on $\T^*(M)$ which is the pullback of differential forms. Then
the canonical $1$-form $\lambda$ on $\T^*(M)$ is invariant for this action and
the symplectic manifold obtained by the hamiltonian reduction $\T^*(M)//G$  with
zero moment level is
canonically
isomorphic to the symplectic space $\T^*(M/G)$ with the canonical symplectic
form.}
\\
{\bf Proof.} Let us describe the cotangent space to $M/G.$ We denote by
$\tau$ the factorization map $\tau:M\rightarrow M/G.$
The tangent space
at the point $m\in M/G$ is by definition
$$
\T_{m}(M/G)=\T_{\tau^{-1}(m)}/\{Im(\mathcal{G})\},
$$
where $Im(\mathcal{G})$ is the subspace in $\T_{\tau^{-1}(m)}$ which corresponds
to the generators  of the action of $G$.
Hence cotangent vectors to  $M/G$ are described by $1$-forms such
that they vanishes on
vector fields coming from the action of the group $G$,
but these are precisely  $1$-forms $\Theta$ such
that the pair $(m,\Theta)$ lies in the zero moment level
because $<\Theta(x)| \xi_g(x)>= <\lambda(x,\Theta)| \tilde \xi_g
(x,\Theta)>= < \mu[(x,\Theta)] | g >$, where
$\tilde \xi_g $ is the canonical lifting of $\xi_g$ from $M$ to $\T^*M$,
$\xi_g$ is the
vector field on $M$ generating the infinitesimal action of the group $G.$

The fact that the canonical symplectic form and the canonical
$1$-form on $\T^*M$ reduce to the
canonical symplectic form and the canonical 1-form on $\T^*(M/G)$ is
quite tautological.
It is enough to note that the
$1$-form $\lambda$ reduced to the zero moment level is correctly defined on
the factor space $\T^*(M/G).$ Indeed, its value on the vector field $\xi$
at the point $m,\theta$ is defined by $<\theta,d\pi(\xi)>$ and does not
depend on the choice of the representative of an element of
$\T_{\tau^{-1}(m)}/\{Im(\mathcal{G})\}~\square$
\vskip 0,5cm
Returning to our specific case we denote the
corresponding symplectic form on $\T^*(GL_r^{\times (N-1)})//{GL_r}$
by $\omega_2.$
Due to proposition \ref{hol-dif-k} we have the identification of spaces
$$\T^*(GL_r^{\times (N-1)})//{GL_r}\stackrel{p}\longrightarrow
\T^*{\M}$$ constructed as follows: for the set of matrices $\{\L_i\}$ one
constructs the bundle $\ML$ and the cotangent vector $\Phi(z)$, which
can be
expressed
from the matrices
$\{\Phi_i\}$ by the usual formula (\ref{hol-dif-k-formula-2}). This
identification is correct by virtue of formula (\ref{mom_map}) for the
moment map.
Different
choices for the representative $\L_i$ correspond to conjugated matrices
$\tilde\L_i=g^{-1}\L_ig,~\tilde\Phi_i=g^{-1}\Phi_ig.$
{\Rem~}
In fact the map $p$ is not only the identification of spaces, it is also
a
symplectomorphism considering the canonical symplectic structures on these
spaces.
Moreover, we will show below that this map  coincides with
the composition of the canonical identification of the spaces
$ \T^*(GL_r^{\times (N-1)})//{GL_r}$  with $
\T^*(GL_r^{\times (N-1)} / {GL_r} )$, described in the lemma above
and of the identification $p$ of $
\T^*(GL_r^{\times (N-1)} / {GL_r} )$  with $\T^*{\M}.$
 The significance of this claim is the following: we reduce the
problem of describing the symplectic form on $\T^*\M$ to the much more
simple question - the symplectic form on the cotangent bundle to the group
(in our case the group is $GL_r$). This non trivial claim in our
description is the definition of the symplectic structure
in Beauville's construction \cite{Bea}, here
we deduce it from the canonical symplectic form on the cotangent bundle.
{\Prop \label{prop_form}
The map $p$ is a symplectomorphism, i.e. $$p^*(\omega)=\omega_2$$
where $\omega$ is the canonical symplectic form on $\T^*\M.$
}
\vskip 0.5cm
As a corollary we see the way to calculate the value of the
canonical $1$-form on the given tangent vector in terms of $\L_i,\Phi_i$:
{\Cor \label{1-form-cor}
Let $\alpha$ be the canonical $1$-form on $\mathcal{T}^* \mathcal{M}$.
Its  value on the vector
$\xi\in\mathcal{T}_{(\Lambda,\Phi)}(\mathcal{T}^*\mathcal{M})$
equals to :
\bea
- \sum_{i=2,...,N} Tr \Phi_i \L_i^{-1} \tilde \xi_i,
\eea
where the point $\Lambda$ in the moduli space of vector bundles
is given by the vector bundle $\ML$, corresponding to the set of matrices
$\L_i$; the covector $\Phi$  corresponds to
$\Phi(z)$ given by formula (\ref{hol-dif-k-formula-2}).
The matrices $\tilde \xi_i$ considered as a tangent vector
to the $(Mat_r^{\times (N-1)})$  at the point
$\L_i, i=2,...,N$ are defined by the condition
that the projection of such a vector onto the space of vector
bundles corresponding to the map $\L_i\mapsto \ML$
coincides with $\pi_{*}(\xi)$, where $\pi$
is the canonical projection from the cotangent bundle to the moduli
space to the moduli space itself.
}
\\
{\bf Proof.} The most simple way to prove this proposition
consists in calculating the
pullback of $\omega$ on $\T^*\M$ to the space
$\mu^{-1}(0)\subset\T^*(\times_{i=2}^N GL_r).$ Let us denote by $\pi_1$ the map
$\pi_1:\mu^{-1}(0)\rightarrow\T^*\M$ which is the composition of the
factorization map and the identification $p.$ Due to the invariance of the
$1$-form $\lambda$ it is sufficient to consider its pullback.  By definition
the value of the ``pullbacked'' $1$-form on a vector $\xi\in\T(\mu^{-1}(0))$ is
$\pi_1^*(\lambda)(\xi)=\lambda(d\pi_1\xi).$ Let us take the vector field
$\xi$ and the deformation $\delta_{\xi}\L_i=i(\xi)d\L_i.$ Further we find $\chi_i$
satisfying the equations:
\begin{equation}\label{chi_po_xi}
\delta_{\xi}\L_i=\chi_1\L_i-\L_i\chi_i
\end{equation}
and chose some $\chi(z)$ such that $\chi(P_i)=\chi_i,~i=1,\ldots,N.$ There
is an ambiguity in equation (\ref{chi_po_xi}), for every $\chi_1$ one
can
find $\chi_i$. In fact the choice of $\chi_1$ does not matter: $\chi(z)$
constructed with different $\chi_1$ lie in the same class
in $H^1(End(E))$ (recall that describing $H^1(End(E))$ in
proposition \ref{lem-chi} we
have factorized the space of matrix polynomials $\chi(z)$
by constant matrices). This map is
really the differential of $\pi_1$ because the tangent vector $\chi(z)$
constructed as said defines the same deformation of $\L_i$ due to proposition
(\ref{chi-deform-lamb}) as the expression (\ref{chi_po_xi}). Now
$$\pi_1^*(\lambda)(\xi)=\lambda(d\pi_1\xi)=<\Phi(z),\chi(z)>$$ where $\Phi(z)$
is constructed from $\{\Phi_i\}$ due to the identification $p.$
In virtue of  Serre's pairing
$$<\Phi(z),\chi(z)>=-\sum_{i=2}^N Tr(\Phi_i\chi(P_i))+
Tr(\sum_{i=2}^N\L_i\Phi_i\L_i^{-1}\chi(P_1)).
$$
On the other hand the canonical $1$-form on $\mu^{-1}(0)$ can be
calculated on the vector $\xi:$
$$\lambda_1(\xi)=\sum_{i=2}^N Tr(\Phi_i \L_i^{-1}\delta_{\xi}\L_i)=
\sum_{i=2}^N Tr(\Phi_i \L_i^{-1}(\chi_1\L_i-\L_i\chi_i))=<\Phi(z),\chi(z)>
$$
$\blacksquare$

\subsubsection{Curves with many cusps}

According to section \ref{end-k-sect},
consider the curve $\Sigma^{proj}$ which is a
$\C P^1 $ curve with $N$ cusps.
Recall that the affine part of $\Sigma^{proj}$ is given by
$ Spec \{ f(z) \in \CC[z]: \forall i=1,...,N   f'(P_i)=0 \}$,
the bundle $\ML$ corresponds to the module which over the affine
part is described as $\{ s(z)\in \CC[z]^r: \forall i=1,...,N ~ s'(P_i)=\L_i s(P_i)
\}$.
In section \ref{end-k-sect} we have considered the case of one cusp,
but everything can be obviously generalized to the case of many cusps.
So the
sections of $H^0(End(\ML)\otimes \K)$ can be described as:
\bea
\Phi (z)= \sum_i \left(\frac{\Phi_i}{(z-P_i)^2} + \frac{[\L_i,\Phi_i] }{z-P_i}
\right)
\eea
where the matrices $\Phi_i$ satisfy
$\sum_{i} [\L_i, \Phi_i ]=0$.
\\
{\bf Claim:}
{\it
The canonical $1$-form on the cotangent bundle to the
 moduli space of vector bundles on the curve $\Sigma^{proj}$
in the coordinates $\L_i,\Phi_i$ can be written as follows:
\bea \label{1-form-cusp}
 \sum_{i=1,...,N } Tr \Phi_i d \L_i
\eea
}
One should slightly correct this formulation as was done in the previous subsection,
which can be done along the same lines; so, we omit all details.

\subsection{Hitchin system}
Let us summarize the results obtained up to now.

{\Th
The Hitchin system on a curve $\Sigma^{proj}$ which is the result of gluing
$N$ distinct points $P_i$ on $\CC P^1$ is the system with phase space
obtained by
hamiltonian reduction from the space of matrices $\L_i,\Phi_i, i=2,...,N$
where $\L_i$ are invertible
with the symplectic form $$\sum_{i=2,...,N} - Tr d(\Phi_i \L_i^{-1}) d \L_i,$$
where $GL(r)$ acts by conjugation.
The Lax operator is given by:
$$\Phi(z)=
\frac{ \sum_{i=2,...,N} -\L_i \Phi_i\L_i^{-1} }{z-P_1}dz
+\sum_{i=2,...,N} \frac{\Phi_i}{z-P_i}dz.$$
}
Its hamiltonians can be obtained by expanding $Tr \Phi(z)^k$
under the basis of holomorphic differentials.
We will give the direct  proof of their commutativity on the nonreduced
phase space
in the more general case in our
next paper \cite{CT2}, so here we omit the details.

Analogously, the Hitchin system on the curve with many cusps
can be described according to:
{\Th
The Hitchin system on a curve $\Sigma^{proj}$ which is the curve $\CC P^1$
with $N$ cusps at distinct points $P_i$
 is the system with phase space obtained by
hamiltonian reduction from the space of matrices $\L_i,\Phi_i, i=2,...,N$,
with the symplectic form $$\sum_{i=2,...,N} - Tr d \Phi_i d \L_i,$$
where $GL(r)$ acts by conjugation.
The Lax operator is given by:
$$
\Phi (z)= \sum_i \left(\frac{\Phi_i}{(z-P_i)^2} +
\frac{[\L_i,\Phi_i] }{z-P_i}\right).
$$
}

\section{Trigonometric and rational Calogero-Moser systems}

\subsection{Node}
As was shown in section \ref{CC} the dualizing sheaf in this case has
one global section $dz(\frac 1 {z-z_1}-\frac 1
{z-z_2}).$ Consider the moduli space $\mathcal{M}$ of holomorphic
bundles $V$ of rank $n$ on $\Sigma_{node}$ with a fixed
trivialization at the point $p,z=z_3.$ It means that
$$\mathcal{T}_V \mathcal{M}=H^1(\Sigma,End(V)\otimes
\mathcal{O}(-p)).$$  We restrict to the principal cell of this
moduli space which corresponds to the space of equivalence classes
of matrices $\Lambda$ with different eigenvalues. The cotangent
space is the space of holomorphic sections of $End^*(V)\otimes
\mathcal{K}\otimes \mathcal{O}(p).$ Such sections are
matrix-valued functions of $z$ of the form $$\Phi(z)=\frac
{\Phi_1}{z-z_1}-\frac {\Phi_2}{z-z_2}+\frac {\Phi_3} {z-z_3}$$
such that
\begin{equation}\label{cond1}
  \Phi_1 \Lambda=\Lambda \Phi_2\quad \mbox{and}\quad \Phi_1-\Phi_2+\Phi_3=0.
\end{equation}
This function is parameterized by $(\Phi_3)_{ij}=f_{ij},i\neq j$,
the eigenvalues $e^{2 x_i}$ of the matrix $\Lambda$ (all calculations are
in the diagonal base for this matrix) and the diagonal elements of
the matrix $(\Phi_1)_{ii}=p_i.$ All other quantities can be
expressed in these terms.

We now investigate the symplectic structure. The moduli space is
parameterized by the matrix $\Lambda$ and the matrix $U$ which fixes
the trivialization at the point $p.$ The
variables $\Phi_i$ define the cotangent vector.
\begin{lem}The canonical symplectic form on the cotangent bundle
$\mathcal{T}^* \mathcal{M}$ can be represented as
\begin{equation}\label{f1}
  \omega=Tr(d (\Lambda^{-1}\Phi_1)\wedge d \Lambda)+
  Tr(d  (U^{-1}\Phi_3) \wedge d  U ).
\end{equation}

\end{lem}
{\Rem ~ \label{sympl-rem} } It is a slightly incorrect formulation:
 we mean that
the canonical symplectic form on the cotangent bundle
$\mathcal{T}^* \mathcal{M}$  can be obtained from that form
after the reduction by conjugation (see section \ref{1-form-sect} for
the precise formulation).
\\
{\bf Proof.} The canonical symplectic form $\omega$ on the
cotangent bundle $\mathcal{T}^* X $ to the manifold $X$ is defined
as follows: the point of the cotangent bundle is the pair $(x,p)$
where $x\in X$ and $p:\mathcal{T}_x X\rightarrow \mathbb{C}.$ We
start by defining the $1$-form $\lambda$ on the cotangent bundle
by the formula
\begin{equation}\label{one}
  \lambda(x,p)(\xi)=p(\pi_* \xi),
\end{equation}
 where
$\xi\in\mathcal{T}(\mathcal{T}^* X),$
 $\pi$ is the projection $\mathcal{T}^* X \rightarrow X$ and
$\pi_*$ is its differential. On the cotangent bundle to
$\mathbb{C}^N$ with coordinates $x_1,\ldots,x_N$ and with the
corresponding coordinates $p_1,\ldots,p_N$ on the cotangent space
this form reads $\lambda=\sum p_i d x_i.$ The canonical symplectic
form on the cotangent bundle is $\omega=d \lambda.$ We have to
prove that the form $\lambda'=Tr(\Lambda^{-1}\Phi_1d \Lambda)+
Tr(U^{-1}\Phi_3 d  U )$ is the canonical $1$-form.

Let us describe explicitly the deformation of the vector bundle $V$
defined by an element $\chi\in H^1(\Sigma,End(V)\otimes
\mathcal{O}(-p)).$ The finite form of this element acts on the
vector bundle data as follows: $$ \Lambda\mapsto
(1+\epsilon\chi(z_1))\Lambda (1-\epsilon\chi(z_2)),\qquad U\mapsto
(1+\epsilon\chi(z_3))U. $$ The infinitesimal form of these
deformations $\delta_\chi
\Lambda=\chi(z_1)\Lambda-\Lambda\chi(z_2)$ and $\delta_\chi
U=\chi(z_3) U $ define vector fields $\xi_\chi$ on the moduli
space and they can be lifted canonically to the vector fields on
the total space of the cotangent bundle. We call this lifting
$l:\mathcal{T}\mathcal{M}\rightarrow\mathcal{T}(\mathcal{T}^*\mathcal{M}),$
and evidently $\pi_* \circ l=id.$ Another type of canonical vector
fields on the total space of the cotangent bundle is the vertical
vector fields $\xi_{vert}$. They act only on the coordinates on
the space of sections $\Phi$ and $\xi_{vert} \in ker \pi_*.$ So we
have to verify that $\lambda'(vert)=0$ and
$\lambda'(l(\xi_\chi))=<\Phi,\chi>.$ The first is trivial because
$d \Lambda(\xi_{vert})=0$ and $dU(\xi_{vert})=0.$ As for the second
condition, one has:
$$\lambda'(l(\xi_\chi))=Tr(\Lambda^{-1}\Phi_1\delta_{\chi}\Lambda)+
Tr(U^{-1}\Phi_3 \delta_{\chi} U )$$
$$=Tr(\Lambda^{-1}\Phi_1(\chi(z_1)\Lambda-\Lambda\chi(z_2)))+
Tr(U^{-1}\Phi_3\chi(z_3)U)$$
$$=Tr(\Phi_1\chi(z_1)-\Phi_2\chi(z_2)+\Phi_3
\chi(z_3))=<\Phi,\chi>,$$ where we have used the paring
(\ref{SerDual}) and the relation $\Phi_1\Lambda=\Lambda\Phi_2.$
$\blacksquare$

To obtain the Calogero-Moser system one needs to perform an additional
hamiltonian reduction.
The form (\ref{f1}) is invariant by the $GL_n$-group
action $g:U\mapsto U g.$ The moment map of this action is $g^{-1}U
g.$ We fix it to be diagonal and factorize the level manifold by
the stabilizer which is $n$-dimensional. This procedure subtracts
$n\times n$ degrees of freedom, corresponding to the auxiliary
variable $U.$ Finally the reduced nonsingular manifold is
$\C^{2n}$ with a canonical nonsingular structure and
the coadjoint orbits of $GL_n$ generically of rank $n$.
This can be
expressed in terms of Poisson brackets as follows:
\begin{equation}\label{br}
  \{x_i,p_j\}=\delta_{ij},\qquad
\{f_{ij},f_{kl}\}=\delta_{jk}f_{il}-\delta_{il}f_{kj}
\end{equation}
{\Rem This expression for the symplectic form can be interpreted
 within
a general approach in terms of coordinates on the nonreduced phase space of
connections. For the constant connection $A$ it is proportional to $\ln\Lambda$
and (\ref{sym}) gives us $\oint
Tr(\delta\Phi\wedge\delta(\ln\Lambda))dz$ which is equal to the
sum of the residues with  the  appropriate signs, i.e. $$\sum_{i=1}^3 Tr(d
\Phi_i\wedge\Lambda^{-1}d \Lambda).$$}

Let us proceed with the construction of integrals. The general scheme gives
the quantities $Tr\Phi^k(z),$ or more precisely all non
trivial negative coefficients of the Laurent expansion of this
expression at the poles. They are in involution by
construction and  by the fact that our symplectic form is obtained
from the canonical one on the cotangent bundle by hamiltonian
reduction. The trace of $\Phi^2(z)$ is a rational function of $z$
the coefficients of which are linear functions of $Tr(\Phi_1^2)$
and $Tr(\Phi_1 \Phi_2).$  A straightforward calculation shows
 that:
\begin{equation}\label{H1}
  H_1=Tr \Phi_1^2= \sum_{i=1}^n p_i^2- 4 \sum_{i\neq j} \frac
  {f_{ij}f_{ji}} {\sinh^2(x_i-x_j)};
\end{equation}
\begin{equation}\label{H2}
  H_2=Tr \Phi_1 \Phi_2 = \sum_{i=1}^n p_i^2 -  \sum_{i\neq j} \frac
  {f_{ij}f_{ji}} {(e^{(2x_j-2x_i)}-1)^2};
\end{equation}
The difference $H_1-H_2$ is equal to the second Casimir element of
$\Phi_3$, i.e. $\Sigma_{i \neq j}f_{ij}f_{ji}.$ Notice that
$H_1$ is the Hamiltonian of the trigonometric Calogero-Moser
system with spin.

\subsection{Cusp}
As in the nodal case we consider the principal cell of the moduli
space which is parameterized by diagonal matrices $\L$ and
introduce the moduli space with fixed trivialization $U$ at the
point $P,z=z_2.$ The section of $End^*(V)\otimes
\mathcal{K}\otimes \mathcal{O}(p)$ in this case can be
represented by $$\Phi(z)=\frac {\Phi_1} {(z-z_1)^2}+\frac {\Phi_2}
{z-z_1}+\frac {\Phi_3} {z-z_2}$$ subject to the conditions
\begin{equation}\label{cond2}
 \Phi_2=[\L,\Phi_1]\quad\mbox{and}\quad\Phi_3=-\Phi_2.
\end{equation}
The coordinates on the phase
space are $x_i$-the diagonal elements of the matrix $\L$, the variables
$p_i$ which are the diagonal elements of the matrix $\Phi_1$,
the nondiagonal elements of the matrix $\Phi_2$ which we denote by
$f_{ij},i\neq j$ and the matrix elements of $U.$ Then solving
the conditions
(\ref{cond2}) one obtains
$${\Phi_1}_{ij}=\frac {f_{ij}} {x_i-x_j}.$$ The infinitesimal
action of the element $\chi\in H^1(\Sigma,End(V)\otimes
\mathcal{O}(-p))$ is the following
$$\delta_{\chi}\L=[\chi(z_1),\L]+\chi'(z_1),\qquad \delta_{\chi}
U=U\chi(z_2).$$ The paring (\ref{SerDual}) realizing Serre's duality
in this case
specializes to the form
$$<\Phi,\chi>=Tr(\Phi_1\chi'(z_1)+\Phi_2\chi(z_1)+\Phi_3\chi(z_2))=$$
$$Tr(\Phi_1(\chi'(z_1)+[\chi(z_1),\L])-\Phi_2\chi(z_2)),$$ where we
used (\ref{cond2}). Substituting the explicit form of the action and
using the arguments from the previous paragraph we obtain:
\begin{lem}The canonical symplectic form on the cotangent bundle
$\mathcal{T}^* \mathcal{M}$ can be represented by
\begin{equation}\label{f2}
  \omega=Tr(d \Phi_1\wedge d  \L)+
  Tr(d  (U^{-1}\Phi_2) \wedge d  U ).
\end{equation}
\end{lem}
{\Rem ~} see remark \ref{sympl-rem}.

As in the previous paragraph we continue by the Hamiltonian
reduction on the variable $U$ and obtain the symplectic variety
corresponding to the brackets (\ref{br}). The coefficients of the
function $Tr \Phi^2(z)$ are linear functions of the expressions
$Tr \Phi_1^2,Tr \Phi_2^2,Tr \Phi_1\Phi_2.$ The first is the
Hamiltonian of the rational Calogero-Moser system with spin
\begin{equation}\label{H3}
  H_3=Tr \Phi_1^2= \sum_{i=1}^n p_i^2+  \sum_{i\neq j} \frac
  {f_{ij}f_{ji}} {(x_i-x_j)^2};
\end{equation}
\begin{equation}\label{H4}
  H_4=Tr \Phi_2^2= \sum_{i\neq j} f_{ij}f_{ji}.
\end{equation}
Let us note that $H_4$ is the second Casimir element for the matrix
$\Phi_2$ and $Tr \Phi_1\Phi_2=0.$

\section{Curves with two cusps}
For smooth curves of genus 2 the moduli space of $SL(2)$-bundles
has been identified in \cite{NR1} with $\CC P^3$.
In the papers \cite{GP,GAW} the Hitchin Hamiltonians
have been written explicitly as functions on $\T^* \CC P^3$.
In our paper we consider singular curves of genus 2
(we consider the example of a rational curve
with only two cusps).
We will describe below the analog of the Narasimhan-Ramanan
parameterization of $SL(2)$ bundles on such a curve
and we will write down the Hitchin hamiltonians.

\subsection{Construction}
In this section we examine the moduli space of $SL_2$ holomorphic
bundles on the curve $\Sigma_2$, defined by the equation
$$y^2=(z-z_1)^3(z-z_2)^3,$$ of algebraic genus $2$ with two
singularities which can be realized by contracting all
cycles on a genus two Riemann surface. On the normalization
which is the rational curve $\mathbb{C}P^1$ obtained
by the following blowup: $t=y/(z-z_1)(z-z_2)$ the inverse image of
the structure sheaf can be realized as the subsheaf
\begin{equation}\label{srt}
  \mathcal{O}_{\Sigma_2}=\{f\in\mathcal{O}_{\mathbb{C}P^1}:\partial_z
f|_{z_1}=\partial_z f|_{z_2}=0 \}.
\end{equation}
The holomorphic bundles $E$ on
$\Sigma_2$ are parameterized by the pairs of matrices $(\L_1,\L_2)$
with zero trace up to common conjugation and can be described in
terms of their section sheaf as follows:
\begin{equation}\label{lin}
 \Gamma_U(E)=\{S\in\mathcal{O}_U(\mathbb{C}P^1)\otimes
\mathbb{C}^2 :\partial_z S=\L_1 S|_{z_1};\partial_z S=\L_2
S|_{z_2}\}.
\end{equation}
For constructing the cotangent bundle $\mathcal{T}^*\mathcal{M}$
to the moduli space of holomorphic bundles $E$ we again exploit
the Kodaira-Spenser correspondence.

Using the same arguments as in previous paragraphs we choose the
realization of the canonical bundle on our singular curve
$\Sigma_2$  to be
$\mathcal{K}(\Sigma_2^{norm})\otimes\mathcal{O}(2z_1+2z_2)$ on
the normalization. The typical section of
$End(V)\otimes\mathcal{K}$ in  this case is the matrix-valued
function $$\Phi(z)=\frac
{\Phi_1}{(z-z_1)^2}+\frac{\Phi_3}{z-z_1}+\frac
{\Phi_2}{(z-z_2)^2}+\frac{\Phi_4}{z-z_2};$$ subject to the
relations:
\begin{equation}\label{cond1}
  \Phi_3=[\L_1,\Phi_1],\qquad\Phi_4=[\L_2,\Phi_2],\qquad
\Phi_3+\Phi_4=0.
\end{equation}

By analogy with the previous section we obtain the following:
\begin{lem}The canonical symplectic form on the cotangent bundle
of $rk=2$ bundles on $\Sigma_2$ can be represented in the form:
\begin{equation}\label{sym1}
  \omega=Tr(d \Phi_1\wedge d  \L_1+d \Phi_2\wedge d  \L_2).
\end{equation}
\end{lem}
{\Rem ~} see remark \ref{sympl-rem}.

As mentioned previously the open subset in the space of holomorphic $rk=2$
bundles in consideration is the space of equivalence classes of
pairs of $2\times2$-matrices $\L_1,\L_2$ up to common conjugation.
The natural coordinates on this space are the invariant functions
$$t_1=Tr\L_1^2,\qquad t_2=Tr\L_1\L_2,\qquad t_3=Tr\L_2^2.$$

Some technical preliminaries are convenient at this point.
Due to the conditions (\ref{cond1}) we have $$
0=-Tr[\L_2,\Phi_2]\L_2 = Tr[\L_1,\Phi_1]\L_2=Tr[\L_2,\L_1]\Phi_1;$$ $$
0=-Tr[\L_1,\Phi_1]\L_1 = Tr[\L_2,\Phi_2]\L_1=Tr[\L_1,\L_2]\Phi_2.$$
Using the fact that the Killing form on $sl_2$ is not degenerate
we conclude that for common $\L_1,\L_2$ the matrices $\Phi_1,\Phi_2$
are linear combinations of them:
\begin{equation}\label{lin1}
  \Phi_i=p_{i1}\L_1+p_{i2}\L_2\qquad i={1,2}.
\end{equation}
Also notice that $p_{12}=p_{21}.$ It comes from
$$p_{12}[\L_1,\L_2]=[\L_1,\Phi_1]=-[\L_2,\Phi_2]=-p_{21}[\L_2,\L_1].$$
Finally we introduce new coordinates:
$$p_1=p_{11}/2;p_2=p_{12}=p_{21};p_3=p_{22}/2.$$
In these coordinates the linear condition (\ref{lin1}) takes  a more convenient
form
$$\Phi_1=2p_1\L_1+p_2\L_2;$$
\begin{equation}\label{lin}
\Phi_2=p_2\L_1+2p_3\L_2.
\end{equation}

\begin{lem} The conjugated variables to $t_1,t_2,t_3$ subject to
the symplectic form (\ref{sym1}) are $p_{1},p_{2},p_{3}$  such that
$$\omega=\sum_{i=1}^3 d p_i \wedge d t_i.$$
\end{lem}
{\bf Proof.} We proceed by comparing the corresponding one-forms
$\lambda=Tr(\Phi_1 d \L_1+\Phi_2 d \L_2)$ and $\lambda'=\sum_{i=1}^3
p_i d t_i.$ The infinitesimal deformation is defined as follows:
$$\delta_{\chi}\L_1=[\chi(z_1),\L_1]+\chi'(z_1);\qquad
\delta_{\chi}\L_2=[\chi(z_2),\L_2]+\chi'(z_2).$$ We have to verify
that $\lambda(l(\xi_\chi))=\lambda'(l(\xi_\chi))$ where $l$ is
some lifting of the vector field on $\mathcal{M}$ to the vector
field on $\mathcal{T}^*\mathcal{M}$ such that $\pi_* l=id.$
$$\lambda(l(\xi_\chi))=Tr(\Phi_1[\chi(z_1),\L_1]+
\phi_1\chi'(z_1)+\phi_2[\chi(z_2),\L_2]+\Phi_2 \chi'(z_2))$$
$$=Tr([\L_1,\phi_1](\chi(z_1)-\chi(x_2))+\phi_1 \chi'(z_1)+\phi_2
\chi'(z_2));$$ where we have used (\ref{cond1}).
$$\lambda'(l(\xi_\chi))=Tr(2p_1
\L_1([\chi(z_1),\L_1]+\chi'(z_1))+p_2\L_1([\chi(z_2),\L_2]$$
$$+\chi'(z_2))+p_2 \L_2([\chi(z_1),\L_1]+\chi'(z_1))+ 2p_3
\L_2([\chi(z_2),\L_2]+\chi'(z_2)))$$ $$=Tr((2p_1 \L_1+p_2
\L_2)\chi'(z_1)+(2p_3\L_2+p_2\L_1)\chi'(z_2)+p_2(\chi(z_1)-\chi(z_2))[\L_1,\L_2]
).$$ Due to the relations (\ref{lin}) one obtains
$[\L_1,\Phi_1]=p_2[\L_1,\L_2]$ which
shows that $\lambda(l(\xi_\chi))=\lambda'(l(\xi_\chi)).$ The fact
that these $1$-forms coincide on the vertical vector fields is
trivial and this ends the demonstration of the lemma $\square$

\subsection{Hitchin Hamiltonians}
To obtain the Hamiltonians we use the natural mapping
$$End(V)\otimes \mathcal{K}\stackrel{\Delta^{\otimes m}}\rightarrow
End(V)^{\otimes^m}\otimes \mathcal{K}^{\otimes^m}\rightarrow
End(V)\otimes \mathcal{K}^{\otimes^m} \stackrel{Tr\otimes id
}\rightarrow \mathcal{K}^{\otimes^m},$$ where the first arrow is the diagonal,
the second is
given by the multiplication in $End(V).$ The composition of these mappings
induces the mapping on cohomologies
\begin{equation}\label{i}
i:H^0(End(E)\otimes\K)\rightarrow H^0 (\K^{\otimes m}).
\end{equation}
Any choice of global sections in $K^{\otimes m}$ gives us a set of
functions on our phase space.

We focus our attention on the
quadratic Hamiltonians which are defined for $m=2.$
Due to the Riemann-Roch theorem
$H^0(\mathcal{K}^2)=3$ and $H^0(\mathcal{K})=2.$ We can take as a basis of
global sections of $\mathcal{K}$ the following
$$s_1=\frac {dz}{(z-z_1)^2};\qquad  s_2=\frac {dz}{(z-z_2)^2}.$$ Their
tensor quadratic monomials
\begin{equation}\label{bas}
s_1\otimes s_1,\quad s_1\otimes s_2
\quad \mbox{and}\quad s_2\otimes s_2
\end{equation}
form a basis in the space
of global sections for $\mathcal{K}^2.$
Now consider the Higgs field
$$\Phi(z)=\left\{\frac
{\Phi_1}{(z-z_1)^2}+\frac {\Phi_2}{(z-z_2)^2}+\frac
{(z_1-z_2)[\L_1,\Phi_1]}{(z-z_1)(z-z_2)}\right\}dz.$$
We calculate the image of (\ref{i})
$$Tr\Phi^2(z)=\left\{\frac
{Tr\Phi_1^2}{(z-z_1)^4}+\frac{Tr\Phi_2^2}{(z-z_2)^4}+\frac{2Tr\Phi_1
\Phi_2+(z_1-z_2)^2 Tr[\L_1,\Phi_1]^2} {(z-z_1)^2(z-z_2)^2}\right\}
dz{\otimes}dz.$$
Decomposing $Tr\Phi(z)^2$ on the basis (\ref{bas}) we obtain the coefficients:
$$H_1=Tr\Phi_1^2=4p_1^2t_1+p_2^2t_3+4p_1p_2t_2;$$
$$H_2=2Tr\Phi_1\Phi_2+(z_1-z_2)^2Tr[\L_1,\Phi_1]^2$$
$$=4p_1p_2t_1+4p_2p_3t_3+(8p_1p_3+2p_2^2)t_2-2(z_1-z_2)^2
p_2^2(t_1t_3-t_2^2);$$
$$H_3=Tr\Phi_2^2=4p_3^2t_3+p_2^2t_1+4p_2p_3t_2;$$ which are the
functions on the original phase space.
{\Prop The quantities $H_i$ are in involution, so one obtains an integrable system.}
The proof is straightforward.
{\Rem Here we have a
one-parameter family of a priori non-equivalent integrable systems on
$\mathcal{T}^*\mathbb{C}^3$. However the identification of this family with
the Neumann model remains unclear.}

\subsection{Degenerated Narasimhan-Ramanan parameterization}
Here we recall the classical construction from \cite{NR1,NR2}
which identifies the moduli space of the stable bundle on a regular
curve $\Sigma$ of genus $2$ with $\mathbb{C}P^3$
$$\mathcal{M}{\stackrel{\Delta}\longrightarrow}|2\Theta|\cong\mathbb{C}P^3.$$
With the bundle $E$ one associates $D_E\subset Pic_1(\Sigma)$ such
that for any line bundles $\mathcal{L}\in D_E$ the dimension of
$H^0(E\otimes\mathcal{L})$ equals $1.$ Due to \cite{NR1,NR2} the
divisor $D_E$ lies in the linear system $|2\Theta|$ and this
correspondence is an isomorphism. $D_E$ is given by the equation
$$\sum_{i,j=1,2}p_{ij}\theta_{[\frac 1 2 (i-1),\frac 1 2
(j-1)]}=0.$$ The coefficients $p_{ij}$ are projective coordinates
on the moduli space.

The degenerate case shares similar considerations.
Let us take
the $SL_2$-bundle $E$ given by the pair of matrices $\L_1,\L_2$ on
$\Sigma_2$ and the linear bundle $\mathcal{L}$ of degree $1$ given
by the pair of complex numbers $\lambda,\mu.$ Their tensor product
$E'=E\otimes \mathcal{L}$ is the $rk=2$ and $deg=2$ holomorphic
bundle. At the singular points this bundle is characterized by the
pair of matrices $\L_1+\lambda \mathbf{1},\L_2+\mu \mathbf{1}.$ We
have to calculate the dimension of $H^0(E\otimes\mathcal{L}).$ To
describe the space of global sections we use the following covering
$U_1=\Sigma_2\backslash\infty$ and
$U_2=U_{\infty,\varepsilon}$-small neighborhood of $\infty.$ The
fact that the bundle $\mathcal{L}$ is of degree $1$ means that the
transition function associated to this covering can be chosen in
the scalar form $z$ which is invertible in $U_1\cap U_2.$ So,
the global sections of $E'$ are linear vector functions $S$ on
$U_1$ such that $\partial_z S=\L_1 S|_{z_1};\partial_z S=\L_2
S|_{z_2}$. Notice that these holomorphic vector functions can be
continued to the chart $U_1$ because they are linear and $S/z$ is
regular at $\infty.$

We rewrite linear defining conditions (\ref{lin}) using that the
section $S$ is linear, i.e. $S(z)=S_0+z S_1$ as follows:
$$S_1=(\L_1+\lambda\mathbf{1})(S_0+z_1S_1);$$
$$S_1=(\L_2+\mu\mathbf{1})(S_0+z_2S_1).$$ The consistency condition
for this linear system is $$Det\left(\begin{array}{cc}
\L_1+\lambda\mathbf{1} & z_1(\L_1+\lambda\mathbf{1})+\mathbf{1}\\
\L_2+\mu\mathbf{1} & z_2(\L_2+\mu\mathbf{1})+\mathbf{1}
\end{array}\right)=0.$$
A straightforward calculation gives the following expression for
the determinant:
$$Det(\lambda,\mu)=\lambda\mu(\lambda(z_1-z_2)-2)(\mu(z_1-z_2)+2)+(\lambda+\mu)^2$$
$$ +(\lambda^2\det(\L_2)+\mu^2\det(\L_1))(z_1-z_2)^2+ 2(\lambda
\det(\L_2)-\mu \det(\L_1))(z_1-z_2)$$
\begin{equation}\label{det}
+\det(\L_1)\det(\L_2)(z_1-z_2)^2+\det(\L_1)+\det(\L_2)+Tr(\L_1\L_2).
\end{equation}
In terms of the variables $t_1,t_2,t_3$ we rewrite (\ref{det}) as:
\begin{equation}\label{det1}
 Det(\lambda,\mu)=\frac {t_1 t_3}{4}(z_1-z_2)^2+t_2 - \frac {t_1}
  2 (\mu (z_1-z_2)-1)^2- \frac {t_3}2(\lambda(z_1-z_2)+1)^2 +(\lambda
  \mu(z_1-z_2)-\lambda+\mu)^2.
\end{equation}
Now we interpret the functions $$1,\quad -
{(\mu(z_1-z_2)-1)^2}/2,\quad -{(\lambda(z_1-z_2)+1)^2}/2,\quad
(\lambda
  \mu(z_1-z_2)+\lambda-\mu)^2 $$ as the basis of $\theta$-functions of second
  order for our singular curve and the expressions
  $$\tau_1=t_1,\quad \tau_2=t_2+\frac {t_1 t_3}{4}(z_1-z_2)^2,\quad \tau_3=t_3$$
as affine analogs of Narasimhan-Ramanan parameters.

\section{Conclusion} We have shown that the invariant description of
Hitchin system on singular curves is consistent and
provides an explicit
parameterization in all considered cases. However, the problem of constructing
nonreduced coordinates remains tricky. We outline below
 several principal directions in which to continue the study of
 Hitchin system:
\begin{itemize}
  \item {\bf Explicit parameterization.} Even in nonsingular cases
there are
some manners to explicitly parameterize the moduli space of (semi-)stable
holomorphic bundles. The Hecke-Turin parameterization is one of the most
universal and the problem here is to reexpress the explicit formulas
obtained above
for the Hamiltonians in terms of the Hecke-Turin parameters
(see \cite{ER11} for the case of nonsingular curves).
For singular curves it will be done in \cite{CST}.
A related question is the
construction of separated variables and the understanding of its algebraic
nature.
  \item {\bf General description of Hitchin system on singular curves.} The
considered examples show the homogeneity of the analysis in such cases as
cusp and node singularities. The subject of the subsequent paper \cite{CT2} is the
universal treatment of a wide class of singular curves.
 \item {\bf Compactification of moduli space of vector bundles on singular
curves.} For the curves considered in this paper (curves with cusps and
nodes) the vector bundles can be described very explicitly - so one hopes
that one can explicitly describe the compactification of the  moduli space of vector
bundles.  The moduli space of linear bundles on  a singular curve
 must be
compactified by torsion free sheaves. So for the case of vector bundles
it is most  likely that it must be compactified by some
"semistable torsion-free sheaves".
  \item {\bf Lax Pair Representation.} All integrable
systems  have a Lax pair representation. It is not unique and for almost all
known integrable systems the Lax
representation arises naturally, as, for example, the auxiliary linear
problem for nonlinear PDE's.
It is important to obtain a Lax pair representation for our system and
to understand its intrinsic meaning.

(The Lax representation of Calogero-Moser system was discussed in the framework
of Hitchin systems in \cite{HurtM}).
  \item {\bf Classical Solutions.} There is a general prescription
to obtain ``action-angle'' variables for Hitchin system
- they are related to the Jacobians of spectral curves.
Another way to solve the classical system is the so-called
projection method which can be specially effective in our description.
It would be interesting to
compare these methods and to obtain explicit solutions of  the classical
equations of motion.
\item{\bf Relation to hierarchies of isomonodromic
deformations and to the Knizhnik-Zamolodchikov-Bernard equations.}
It is known (see for example \cite{LO}) that  by changing the complex structure
on  a curve one goes from a Hitchin
system to  hierarchies of isomonodromic deformations and from a
quantum Hitchin system to the Knizhnik-Zamolodchikov-Bernard equations.
It would be tempting to consider the movement of double points and cusps
and to obtain such equations explicitly in our case.
  \item {\bf Quantization.}
Another  interesting goal is
to find explicitly the quantum commuting hamiltonians
related to  the classical ones presented here.
One hopes to fully explore the Hitchin system: to find the
wave-functions, the spectrum,  the statistical sum,  the correlators, to explore
the so-called duality \cite{FGNR,GR} for Hitchin system.
Also it would be interesting to  explore  the properties of wave-functions:
their integral representations, asymptotics, different relations
among them and so on.
The quantization of Hitchin
system goes back to \cite{BD} and was discussed later
for some particular
cases in \cite{FFR,Fr,ER11}.  The rather general approach of
\cite{BT} provides
the quantization of  an analog of the Hitchin system, namely a system over the
space of rational matrices (the Beauville system),
in terms of separated variables.
The quantization of the Hitchin and related Beauville-Mukai
systems was discussed recently in \cite{ER2,ER3}.
Another approach  to quantization
is built on the notion of quantum R-matrix and its generalization
\cite{ABB}  (let us note that the classical r-matrices
for the Hitchin system were found in \cite{ER1,VALD}).
It would be very interesting to understand the analogs of all  these
constructions for Hitchin systems on singular curves.

  \item {\bf Separation of variables and geometric Langlands correspondence.}
One of the most interesting goals is to work out explicitly the separation
of variables (\cite{Skl,BT}) for the integrable systems
considered in our paper and to understand its
relation to the geometric Langlands correspondence (\cite{BD,Fr,EFR,FGNR}).
It seems that the integrable systems considered here are quite simple and
explicit so one can try to explicitly understand some complicated
constructions of \cite{BD} in these cases and to shed some light on the
miracle of the Langlands correspondence.
One  must also note that the curves considered in this paper
have been considered in \cite{Serre} for the
construction of ramified geometric class field theories;
one can hope that these singular curves  will play  an analogous role
for the ramified version of the geometric Langlands correspondence.

\end{itemize}


\end{document}